\documentclass[preprint,3p,times]{elsarticle}

\journal{Radiation Physics and Chemistry}

\usepackage{graphicx}
\usepackage{rotating}
\usepackage{amssymb}
\usepackage{mathptmx}
\usepackage{enumitem}
\usepackage{amsmath}
\usepackage{bm}
\usepackage{natbib}

\makeatletter

\newcommand{\dif}    {\ensuremath{\,\mathrm{d}}} 

\newcommand{\be}{\begin{equation}}
\newcommand{\ee}{\end{equation}}
\newcommand{\ba}{\begin{align}}
\newcommand{\ea}{\end{align}}

\usepackage{xcolor}
\definecolor{darkgreen}{rgb}{0.1,0.6,0.1}
\definecolor{steelblue}{rgb}{.273,.508,.703}

\newcommand{\alert}[1]{{\color{darkgreen}#1}}

\newcommand{\removed}[1]{{\color{midnight}\sout{#1}}}

\renewcommand{\removed}[1]{}

\usepackage{hyperref}
\hypersetup{
    colorlinks=true,
    pdftitle={RFI\_lanthanide\_L\_lines},
    pdfauthor={Joseph Fowler},
    pdfpagemode=FullScreen,
}

\begin{document}

\begin{frontmatter}
\author[nist,cu]{J.W.~Fowler\corref{cor1}} \ead{joe.fowler@nist.gov}
\author[nist686]{L.~Miaja-Avila}
\author[nist]{G.~C.~O'Neil}
\author[nist,cu]{J.~N.~Ullom}
\author[cu]{H.~Whitelock}
\author[nist]{D.~S.~Swetz}

\cortext[cor1]{Corresponding author}
\affiliation[nist]{
    organization={National Institute of Standards and Technology, Quantum Electromagnetics Division},
    addressline={325 Broadway},
    city={Boulder}, 
    state={Colorado},
    postcode={80305},
    country={USA}
}
\affiliation[cu]{
    organization={University of Colorado, Department of Physics},
    city={Boulder}, 
    state={Colorado},
    postcode={80309},
    country={USA}
}
\affiliation[nist686]{
    organization={National Institute of Standards and Technology, Applied Physics Division},
    addressline={325 Broadway},
    city={Boulder}, 
    state={Colorado},
    postcode={80305},
    country={USA}
}
\title{The potential of microcalorimeter x-ray spectrometers for measurement of relative fluorescence-line intensities}

\begin{abstract}
We have previously used an array of cryogenic microcalorimeters with 4 eV energy resolution to measure emission-line profiles and energies of the characteristic L-shell x~rays of four elements of the lanthanide series: praseodymium, neodymium, terbium, and holmium. We consider the power of the same data set for the estimation of the lines' relative intensities. Intensities must be corrected for detector efficiency and self-absorption, and we estimate uncertainties on the corrections. These data represent one of the first uses of cryogenic energy-dispersive sensors to estimate the relative intensities of x-ray fluorescence lines. They show that a future measurement of thin-film samples with microcalorimeter detectors could achieve systematic uncertainties below 1\,\% on relative line intensities over a broad energy range.

\end{abstract}

\end{frontmatter}


\section{Introduction}

X-ray fluorescence (XRF) radiation is the result of excited atoms emitting x~rays at several specific energies characteristic of the element, as outer-shell electrons fill inner-shell vacancies. The XRF spectrum serves as a kind of ``atomic fingerprint'' from which the composition of a complex sample can be determined. Quantitative materials analysis is possible, either through comparison with reference samples of known composition, or through the reference-free \emph{fundamental parameters} (FP) method. Users of the FP method model excitation and emission processes and the passage of x~rays through matter.

An accurate emission model requires knowledge of line energies and widths (or even full line profiles, when a spectrum will be measured with high energy resolution). It also requires the relative intensities of each detectable emission line. The success of the FP method depends on tabulations of atomic data that are both accurate and complete. While some FP values are well established, many are not.

As one unhappy example among many, there exists no complete, experimental tabulation of relative intensities for the numerous L lines of rare-earth elements in the 5\,keV to 15\,keV range. This knowledge gap results from the challenges of measuring spectra over a wide range of both energies and intensities with an instrumental response that is well-calibrated and stable. Some work half a century old used diffractometers with $\sim20\,$eV resolution to measure the more intense lines of selected rare-earth elements~\cite{Salem1971,McCrary1972}. Measurements have more commonly and more recently been made with Si(Li) or silicon drift detectors with modest energy resolution of approximately 150\,eV, which is inadequate to resolve many prominent lines. Therefore, authors frequently estimate the combined intensity of a family of lines such as the L$\beta$ lines; unfortunately, the L$\beta$ family blends emission from the L1, L2, and L3 subshells---whose excitation can vary independently---into a single intensity. There exist several measurements of L$\beta$/L$\alpha$, L$\gamma$/L$\alpha$, or L$\ell$/L$\alpha$ ratios for one or more of the elements studied in the present work \cite{Raghavaiah1987, Raghavaiah1990, Rao1993,Barrea2000,Durak2001, Oz2004,Salah2005,Yalcn2008,Alqadi2013} and for other rare-earth elements~\cite{Garg1984,Rao1995,Ismail2000,Gurol2003,Demir2008,Durdu2012,Wang2015,Ganly2016}. The dependence of line energies, line widths, and intensity ratios on chemical state have been studied for some lanthanide-series elements~\cite{Durdag2017}. Recent measurements of intensity ratios for several well-resolved L~lines have been made for bismuth with a solid-state detector~\cite{Menesguen2018}, for gadolinium with a high-resolution von Hamos spectrometer~\cite{Wansleben2019}, and for three actinide elements with a cryogenic microcalorimeter~\cite{Mariam2022}.

In this work, we estimate the \emph{relative fluorescence intensity} (RFI) of many L lines of lanthanide elements. We define the RFI of an emission line as a ratio: the number of x-ray photons emitted in that line, divided by those from the most intense emission line in the same series. The RFI of any line in the L3 series, for instance, is its intensity relative to the L$\alpha_1$ reference line.\footnote{Some other authors define RFI as a fraction instead: the number of photons from a given line divided by those from all lines in a series, such as the series of all emission lines due to L3-subshell vacancies. Section~\ref{sec:results} discusses why we favor the definition based on reference lines.} Ratios that mix more than one subshell, such as the L$\beta$/L$\alpha$ ratio, are less fundamental and are not estimated here. Auger transitions, with or without radiative emission, play no role.
 
RFI values appear in some x-ray FP tabulations based on theoretical calculations, with interpolation of values for those elements lacking complete theoretical models. Calculations by Salem~\cite{Salem1974} are the basis for a modern interpolation by Elam~\cite{Elam2002}, available in the {\tt xraydb} Python library\footnote{\url{https://xraypy.github.io/XrayDB/}}. Calculations by Scofield~\cite{Scofield1974} are the basis of the {\tt xraylib} library\footnote{\url{http://lvserver.ugent.be/xraylib-web/} 
has bindings for several programming languages}~\cite{Schoonjans2011}. The reliability of the RFI data in such databases is difficult to assess, though their completeness is an indisputable advantage.

Our recent measurement has established that cryogenic x-ray microcalorimeters offer several benefits for the metrological characterization of the energies and profiles of fluorescence lines~\cite{Fowler2021}. Microcalorimeters can have energy-resolving power close to that of wavelength-dispersive spectrometers, but with large advantages in collection efficiency and the ability to measure a wide simultaneous energy band~\cite{Ullom2015}. In this work we explore the question of whether they can also be used for estimation of relative intensities, and what systematic uncertainties might limit such a measurement. We estimate the RFI of some L lines of four metals from the lanthanide series. We report the uncertainties that arise from existing data taken with an instrument not optimized for RFI estimation, as well as the uncertainties that could be achieved in a targeted, future measurement. Our goal is the measurement of x-ray fundamental parameters of many elements to support modeling of complete fluorescence spectra.

\section{Experimental method}

This analysis is based on x-ray emission spectra between 5\,keV and 10\,keV measured with superconducting cryogenic microcalorimeters~\cite{Doriese:2017}. We summarize the spectrometer, measurement, and energy calibration in this section; a much more detailed description appears with appropriate references in an earlier publication~\cite{Fowler2021}.

An array of 192 transition-edge sensors (TESs) was operated with only 64 sensors active to minimize cross-talk effects that can distort the x-ray spectra. Each TES employs an absorbing layer of gold 1\,$\mu$m thick to thermalize the energy carried by an x-ray photon. A bilayer of molybdenum and copper with a superconducting transition temperature of 111\,mK acts as a resistive thermometer; it converts the temperature change caused by an absorbed photon into a transient pulse in the bias current. Operation at such low temperatures reduces thermal noise to the point that single photons with energies up to approximately 11\,keV are measured with an energy resolution of 4\,eV (Gaussian FWHM). Weak thermal contact with a cold 70\,mK bath and negative electrothermal feedback in the bias circuit work together to return the TES to its quiescent state within milliseconds of an x-ray detection. A time-division multiplexing system~\cite{Doriese:2016}  reads out the 64 TESs through only eight SQUID-based amplifier chains.

The excitation source was a commercial x-ray tube that accelerated electrons through 12.5\,kV onto a primary target of tungsten. X~rays produced in the tungsten (mostly bremsstrahlung, but also W fluorescence) passed through an aperture approximately 4\,mm in diameter to excite the secondary target, which contained the fluorescence samples of interest. Multiple samples were mounted on a rotary sample holder and alternated at least once every 60 seconds to ensure that variations in sensor gain over time scales of hours could be adequately monitored and corrected. The holder is aluminum, so any small areas not covered by samples emit fluorescence below the energy range of interest. Half the samples in each measurement were foils of rare-earth metals from the lanthanide series: Pr, Nd, Tb, and Ho ($Z=59,60,65$, and 67). These metals were sourced from a chemical supply company. They were 99.9\% pure except for the praseodymium foil, which had a 0.4\% neodymium content according to the supplier’s assay. The other samples were mixed foils of multiple metals from the 3d transition elements. The 3d metals have well-characterized K~lines in the 5\,keV to 10\,keV x-ray band~\cite{Holzer1997,Mendenhall2017,Chantler2006,Chantler2013}, so their fluorescence emission was used to construct an absolute calibration connecting x-ray pulse amplitudes with the corresponding photon energies. The transition-metal samples also allowed us to measure the sensors' energy resolution, to verify their Gaussian energy response, and to bound sources of systematic uncertainty such as time-varying energy calibration. The full measurement of both lanthanide and calibration samples used 107 hours of data collected over a ten-day period.

A series of materials located between the emission samples and the TES array affected the spectrometer's efficiency for photons of various energies. Fluorescent x~rays passed through 7\,cm of air, partially evacuated to an absolute pressure of roughly 40 kPa. A commercially sourced x-ray vacuum window separated the sample space from the cryogenically refrigerated volume that contained the TES array. The window consisted of an aluminized polymer film, backed by a thick stainless steel mesh (filling factor: $19\,\%\pm1\,\%$). A series of aluminum foils blocked infrared radiation in the cryogenic space, with a total foil thickness of 22.7\,$\mu$m. The overall efficiency of the spectrometer and its uncertainty are quantified in Section~\ref{sec:efficiency}.

The energy-calibration procedure used several intense emission lines with published, absolutely calibrated peak energies and line profile shapes as ``anchor points''. Most anchor points were the K$\alpha$ and K$\beta$ emission of 3d transition metals (Ti, V, Cr, Mn, Fe, Co, Ni, and Cu), but the Si K$\alpha$ peak and several L lines of W were also used in order to calibrate the widest possible energy range. A distinct calibration curve from detected pulse amplitude to energy was created for each TES detector~\cite{Fowler2022}. Cross-validation tests in which one or two anchor points were temporarily omitted from the calibration established that the calibration uncertainty is less than 0.2\,eV from 5\,keV to 7.5\,keV and less than 0.5\,eV for all L lines of the elements we studied~\cite{Fowler2021}.

\section{Analysis: unbiased estimates of relative intensities}

\begin{figure}[th]
    \centering
    \includegraphics[width=\textwidth]{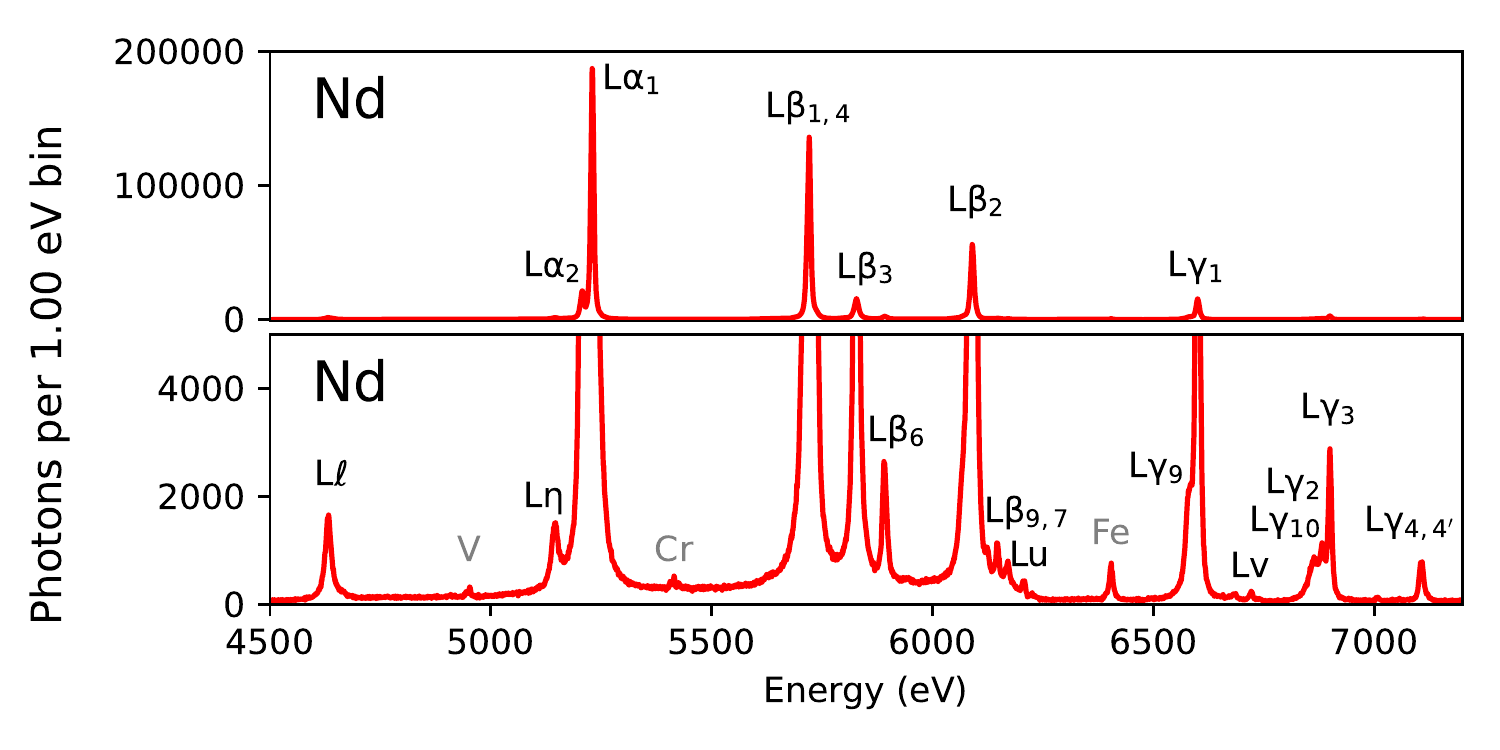}
    \caption{Neodymium fluorescence spectrum from which the relative intensity results derive, shown twice at different y-axis scales. The spectra shown here are not corrected for detector or self-absorption efficiencies. Low-intensity peaks marked V, Cr, or Fe are the K$\alpha$ emission of the indicated 3d metals.}
    \label{fig:nd_spectrum}
\end{figure}

To estimate the emitted relative fluorescence intensities from an energy-calibrated spectrum like the example in Figure~\ref{fig:nd_spectrum}, we must correct for two measurement biases. One is the variation of the spectrometer's detection efficiency with energy. The other bias arises from absorption of fluorescent emission within the emitting sample, or \emph{self-absorption}; it depends on the fluorescence energy and on which L~subshell has a vacancy filled by the emission line in question.  Self-absorption could be minimized by the use of optically thin samples, at the cost of much reduced signal. We did not use thin samples in the current work---all foils were at least 250\,$\mu$m thick. In this section, we estimate the correction factors and the uncertainties in them. All uncertainties specified in this work are standard (1-$\sigma$) uncertainties.

\subsection{Estimation of measured relative intensities from the emission spectra}

The detected relative intensities are extracted from the broad-band emission spectra as part of the spectral-modeling procedure. Each spectral region of interest (ROI) contains one or more emission features: individual lines, doublets, or more complex unresolved features. The intensity estimation is complicated by the fact that some lines overlap one another, and some have very low intensities relative to the background level. We fit each ROI as a model of the background (described in the next paragraph), plus the product of the detector efficiency and the sum of one or more Voigt functions in energy. A Voigt function is the convolution of a Gaussian and a Lorentzian. We intend the Lorentzian to represent the intrinsic, long-tailed lineshape of any single atomic-emission channel and the Gaussian to represent both instrumental broadening and an unresolved population of very similar initial-final state energy differences. Details and the best-fit line models appear in our earlier work~\cite{Fowler2021}.

The background model is the sum of three components. In each ROI, a straight line with two non-negative free parameters (the level at either end of the ROI) is used to model background due to bremsstrahlung and scattering. The relatively narrow energy range of each ROI means that curvature in the background spectrum need not be included. The other two background components are fit globally, so they have no free parameters in any single ROI\@. One is the gold M-escape effect, in which a single $\sim 2$\,keV characteristic gold M$\alpha$ or M$\beta$ photon escapes without detection from the TES's gold absorbing layer. We account for the energy-dependence of gold escape intensities by a fit to a model with a linear dependence on energy. The escape effect produces two faint echoes of the true spectrum at lower intensity and lower energy, resulting in a small number of discernible escape peaks. The other background is the emission of trace elements, including trace components of the rare-earth or calibration samples and K lines of certain 3d transition metals such as Cr, Fe, Ni, and Cu, which are found in the apparatus near the detector. The most intense L lines of interest in this study are at least 200 times more intense than any of the trace-element emission lines or escape peaks; furthermore, the background peaks coincide with only a few ROIs~\cite{Fowler2021} and affect very few lines.

Overlapping features in  spectrum (such as the Nd L$\gamma_{10,2,3}$ triplet near 6900\,eV in Figure~\ref{fig:nd_spectrum}) can be difficult to disentangle. Allocation of the detected photons among two or more ``lines'' inevitably requires subjective judgements. We have included a systematic-uncertainty term for many overlapping lines to account for this ambiguity as well as possible, but the true line-separation uncertainty is very difficult to estimate.

The number of photons detected was approximately $4\times 10^6$, $2\times 10^6$, and $4\times 10^5$ per element for the L3, L2, and L1 families of lines. At least 20 lines were identified for each element, most with statistical uncertainty on the intensity of 1\,\% or better.

\subsection{Spectrometer detection efficiency} \label{sec:efficiency}

\begin{figure}[th]
    \centering
    \includegraphics[width=\textwidth]{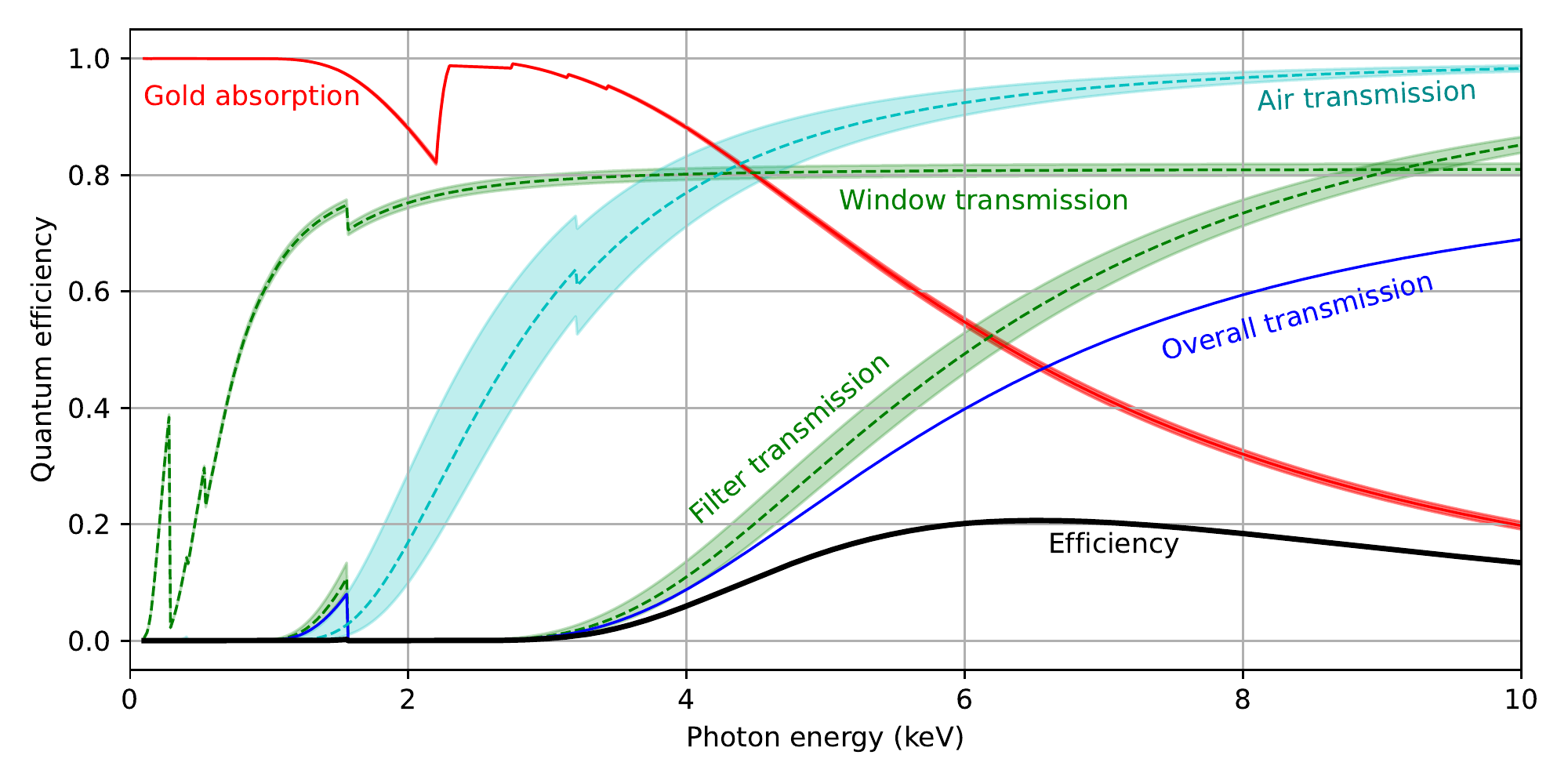}
    \caption{Detector efficiency model. The gold absorber on each TES is 966$\pm20$\,nm thick. The three IR-blocking filters of aluminum are 22.7$\pm1.3$\,$\mu$m thick in total.
    \emph{Air transmission} refers to the passage through 7\,cm of air at $0.40\pm0.12$\,atm pressure. The polymer vacuum window is supported by a grid of stainless steel with clear area factor of 0.81$\pm0.01$. All uncertainties are indicated by shaded bands. For relative intensity measurements, only the uncertainty on the slopes of the curves is relevant. The \emph{Overall transmission} is the product of air, window, and filter transmission. The curve \emph{Efficiency} is the product of transmission and the gold absorption; the peak efficiency is 20.7\,\% at 6.5\,keV\@.
    }
    \label{fig:efficiency}
\end{figure}

\begin{figure}[th]
    \centering
    \includegraphics[width=\textwidth]{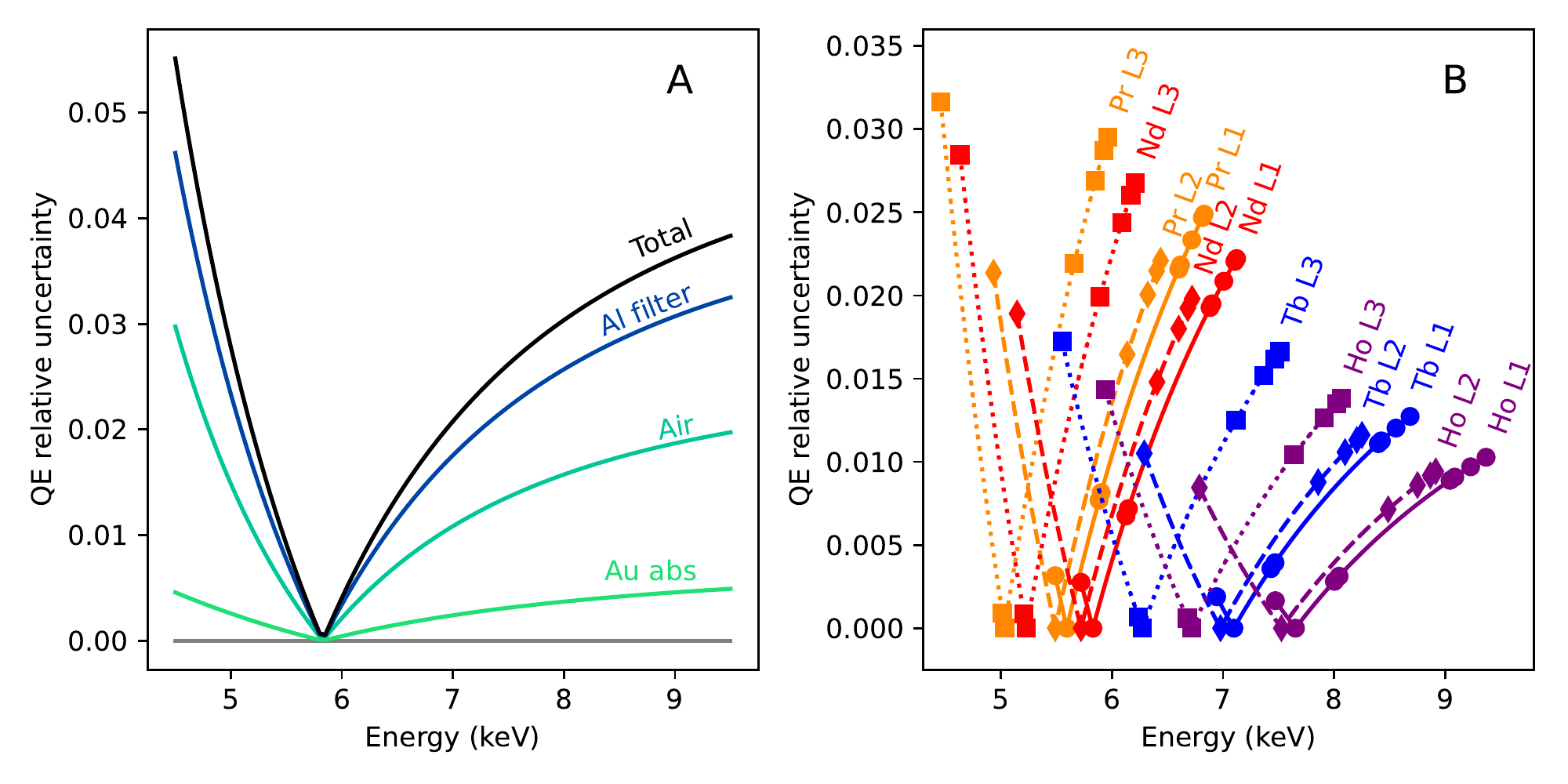}
    \caption{Uncertainties on the detector efficiency. 
    (A) Major factors that contribute to uncertainty on the efficiency, relative to one example reference energy (5.828\,keV, the Nd L$\beta_3$ line). \emph{Al filter} represents $\pm1.3\,\mu$m uncertainty on the aluminum filter thickness; \emph{Air} represents $\pm30\,\%$ uncertainty on the air pressure of 0.4\,atm; \emph{Au abs} represents $\pm20$\,nm ($\pm2\,\%$) uncertainty on the thickness (or surface density) of the gold absorber. \emph{Total} is the quadrature sum of these.
    (B) The total relative uncertainty as shown in Panel A, but here referenced to the energy of the most intense x-ray emission line from each L subshell of each element. The detected emission lines are shown as markers for the L3 ($\blacksquare$), L2 ($\blacklozenge$), and L1 ($\bullet$) subshells. 
    }
    \label{fig:efficiency_uncert}
\end{figure}

The spectrometer's detection efficiency (DE) is computed from the x-ray mass-attenuation properties of all transmitting and absorbing elements in the x~rays' optical path, assuming normal incidence.\footnote{Package {\tt xraydb} was used for mass-attenuation data.} The efficiency model includes the partial x-ray transmission of the vacuum window and three infrared-blocking aluminum filters, as well as the incomplete absorption of photons in the detectors. Transmission and absorption effects set lower and upper limits, respectively, on the usable energy range of the spectrometer (Figure~\ref{fig:efficiency}). The peak efficiency is 20.7\,\% at 6.5\,keV and exceeds 10\,\% over the range 4.5\,keV to 21\,keV\@.

In measurements of relative intensity, results are unaffected by any overall scale factor in the DE model; systematic uncertainties on the RFI arise only from uncertainty in how efficiency varies with energy---slope uncertainties, to leading order. Relative to any chosen reference energy, uncertainties grow with energy difference from that reference (Figure~\ref{fig:efficiency_uncert}A). Factors that contribute to the slope uncertainty of DE include the air pressure in the sample volume ($\pm30\,\%$ uncertain) and the $\pm1.3\,\mu$m uncertainty on the thickness of the aluminum, IR-blocking filters.\footnote{These two uncertainties will be made smaller for future measurements, now that we know they can dominate the relative-intensity uncertainties.} Smaller effects include the uncertain mass-attenuation coefficients of the filters and of the gold absorber. We model these values as uncertain by $\pm1\,\%$ and with slopes known to the level of $\pm0.3\,\%$ per keV, which yield RFI uncertainties up to 1\%. The gold absorber has an uncertainty in the surface density of $\pm2\,\%$, equivalent to a thickness uncertainty of $\pm20$\,nm assuming the density is equal to that of bulk gold (as expected for films thicker than $\sim100$\,nm~\cite{Lovell1968,Siegel2011}). The absorber effect on the RFI results is  $<0.2\,\%$. Other uncertainties were found to affect the relative intensities by much less than 0.1\,\%, including the vacuum window's parameters (fill factor and thickness of the supporting mesh, surface density of the polymer) and the probability of photon escape from the gold absorber. Data-quality cuts also have minimal energy dependence in the relevant energy range. Within a single subshell, relative to the most intense line from that subshell, uncertainties in the DE correction reach 3\,\% for certain lines of Pr and Nd; they are less than 2\,\% for all lines of Tb and Ho (Figure~\ref{fig:efficiency_uncert}B).

\subsection{Self-absorption corrections} \label{sec:self-absorb}


The larger correction to our RFI measurements is for absorption within the sample itself. Any sample thicker than a monatomic layer emits fluorescent x~rays at a range of depths. Emission depths are typically a few microns below the surface in the case of excitation by $\sim 10$\,keV x~rays. Because these depths are similar to the x-ray absorption length, an appreciable fraction of the emission is absorbed in the sample. This fraction depends upon the emitted energy, so relative line intensities must be corrected for the effect. Self-absorption (SA) can be modeled given the excitation spectrum, measurement geometry, and x-ray properties of the sample metals. Like efficiency, the SA correction affects relative intensities only through its energy dependence; the absolute amount of the self-absorption does not affect the RFI results.

\begin{figure}[th]
    \centering
    \includegraphics[width=\textwidth]{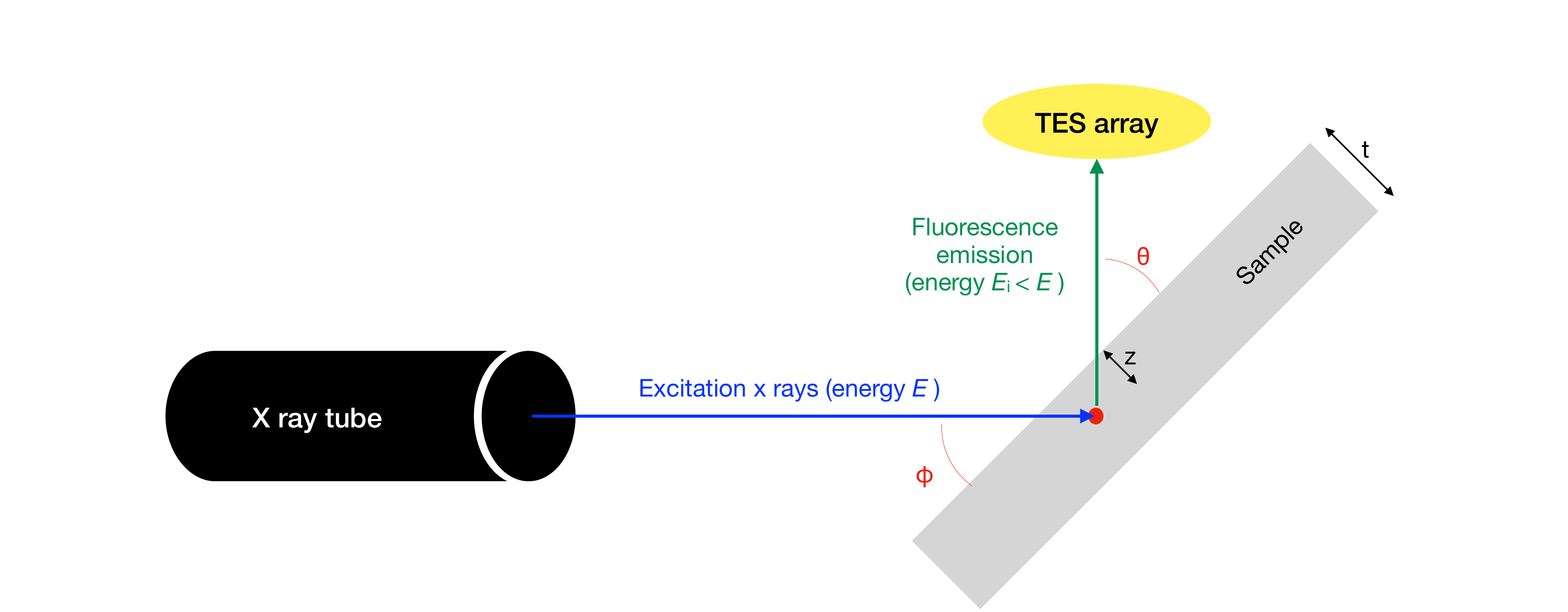}
    \caption{Measurement geometry. An x-ray photon of energy $E$ from the tube source strikes the rare-earth or calibration sample of thickness $t$ at an angle $\phi$ to its surface. A fluorescent photon of energy $E_i$ is emitted from a depth $z$ in the sample at an angle $\theta$ to the surface and is collected at the TES array. Its probability of emerging from the sample, when averaged over all possible depths $z\le t$ and excitation energies $E$, is the self-absorption correction $A_i$ for emission line $i$.}
    \label{fig:geometry}
\end{figure}

We define the SA correction $A_i$ as the probability that an emitted x-ray photon of fluorescence line $i$ and energy $E_i$ will emerge from the rare-earth sample without being absorbed. Let $x$ be the L-subshell filled by the line ($x \in \{1,2,3\}$). The probability $A_i$ must be averaged over the appropriate distributions of emission depths and---because we use a broad-band excitation source---excitation energies. The linear x-ray attenuation coefficient $\mu(E)$ is a function of energy, with units cm$^{-1}$. For any specific emission depth $z$ and an angle $\theta$ between the sample surface and the detector (Figure~\ref{fig:geometry}), this survival probability is  $A_i(z)=\exp[-\mu(E_i)z/\sin\theta$]. Although the TES elements at the edge of the array can differ from the central value of $\theta$ by $\pm 3^\circ$, the range of $\theta$ has negligible effect on the relative intensity estimates.

The depth $z$ is not fixed. It is exponentially distributed for a given excitation energy $E$, with mean depth $\langle z \rangle=\sin\phi/\mu(E)$. The normalized depth distribution is
\begin{equation} \label{eq:Pz}
    P_z = [\mu(E)/\sin\phi]\ \exp[-\mu(E)z/\sin\phi].
\end{equation}

The distribution of excitation energies $P_{E,x}$ that cause an L$x$ vacancy is proportional to the photon spectrum $S(E)$ of the excitation source and to a weighting factor $\mu_x(E)/\mu(E)<1$, the fraction of all interactions that produce a vacancy in L$x$ subshell. Necessarily, $\mu_x(E)=0$ when $E<E_x$, the edge energy of the L$x$ subshell. Thus
\begin{equation} \label{eq:PE}
    P_{E,x} =N\frac{\mu_x(E)}{\mu(E)}S(E).
\end{equation}
$N$ is the normalization factor (computed numerically) that ensures $1=\int_{E_x}^\infty\,\dif E P_{E,x}$.

Equation~\ref{eq:PE} is incomplete for L-shell fluorescence. We must also account for Coster-Kronig (CK) transitions, in which an electron from a higher L subshell spontaneously fills a vacancy in a lower L subshell~\cite{Coster1935}. CK transitions increase the effective rate of L3 and L2 vacancies by creating them out of what were initially L2 or L1 vacancies. We correct Equation~\ref{eq:PE} with the use of the effective $\widetilde{\mu}_x$ instead of $\mu_x$:
\begin{align*}
    \widetilde{\mu}_1(E) &= \mu_1(E) \\
    \widetilde{\mu}_2(E) &= \mu_2(E) + f_{12}\mu_1(E)\\
    \widetilde{\mu}_3(E) &= \mu_3(E) + f_{23}\mu_2(E) + (f_{13}+f_{12}f_{23})\mu_1(E)
\end{align*}
where $f_{jk}$ are the element's Coster-Kronig factors for electronic transitions from the L$k$ to the L$j$ subshell.

If we define a geometrical extinction rate
\begin{equation*}
    \chi(E,E_i) \equiv \frac{\mu(E)}{\sin\phi}+\frac{\mu(E_i)}{\sin\theta}
\end{equation*}
and use the probability distributions for $z$ and $E$ (Equations~\ref{eq:Pz} and \ref{eq:PE}), the differential survival probability is
\begin{align*}
    \dif A_i &= \exp[-\mu(E_i)z/\sin\theta]\ P_z(z,E)\ P_{E,x}(E)\ \dif z\ \dif E \\
    &= \exp[-\chi(E,E_i) z]\ \frac{\mu(E)}{\sin\phi}\ N\ \frac{\widetilde{\mu}_x(E)}{\mu(E)}\ S(E)\ \dif z\ \dif E.
\end{align*}
We can perform the $z$ integral over the range $[0,t]$ analytically. The exponential factor integrates to $(1-\exp[-t\chi])/\chi$. The samples used in this work all have thickness $t\ge250\,\mu$m; for such samples, $t\chi\gg 1$, in which limit the integral becomes $1/\chi$. The thick-sample correction is thus
\begin{align}
    A_i  &= \int_{E_x}^{E_\mathrm{max}}\,\dif E\  \frac{1}{\chi(E,E_i)}\frac{\mu(E)}{\sin\phi}\ N\ \frac{\widetilde{\mu}_x(E)}{\mu(E)}\ S(E)  \nonumber \\
    &= \left .
    \int_{E_x}^{E_\mathrm{max}}\,\dif E\  \left(1+\frac{\sin\phi}{\sin\theta}\frac{\mu(E_i)}{\mu(E)} \right)^{-1} \frac{\widetilde{\mu}_x(E)}{\mu(E)} S(E) 
    \middle/
    \int_{E_x}^{E_\mathrm{max}}\,\dif E\  
    \frac{\widetilde{\mu}_x(E)}{\mu(E)} S(E),
    \right. \label{eq:sa_weights}
\end{align}
where $E_\mathrm{max}$ is the maximum energy in the excitation spectrum, and Equation~\ref{eq:sa_weights} gives the normalization factor $N$ explicitly. Thus $A_i$ is an average over excitation energies of $[1+\sin\phi\mu(E_i)/(\sin\theta\mu(E))]^{-1}$, weighted by the product of the excitation spectrum $S(E)$ and the effective photoionization fraction (i.e., the fraction of all interactions that produce a vacancy in subshell L$x$, with CK transitions considered).

Measurements of the excitation spectrum show that $S(E)$ is consistent with a simple model of bremsstrahlung emission up to a cutoff energy of $E_\mathrm{max}=12.5$\,keV, plus a small addition of the most intense L lines of the primary electron target, tungsten:
\[
S(E) \propto \left(\frac{E_\mathrm{max}}{E}-1\right) + S_\mathrm{WL}.
\]
The form of $S(E)$ below the L3 edge (and specifically its low-energy cutoff) need not be modeled, because $\widetilde{\mu}_x(E)=0$ for energies below the L$x$ absorption edge, $E_x$ (6.0\,keV for Pr L3). 

\begin{figure}[th]
    \centering
    \includegraphics[width=.94\textwidth]{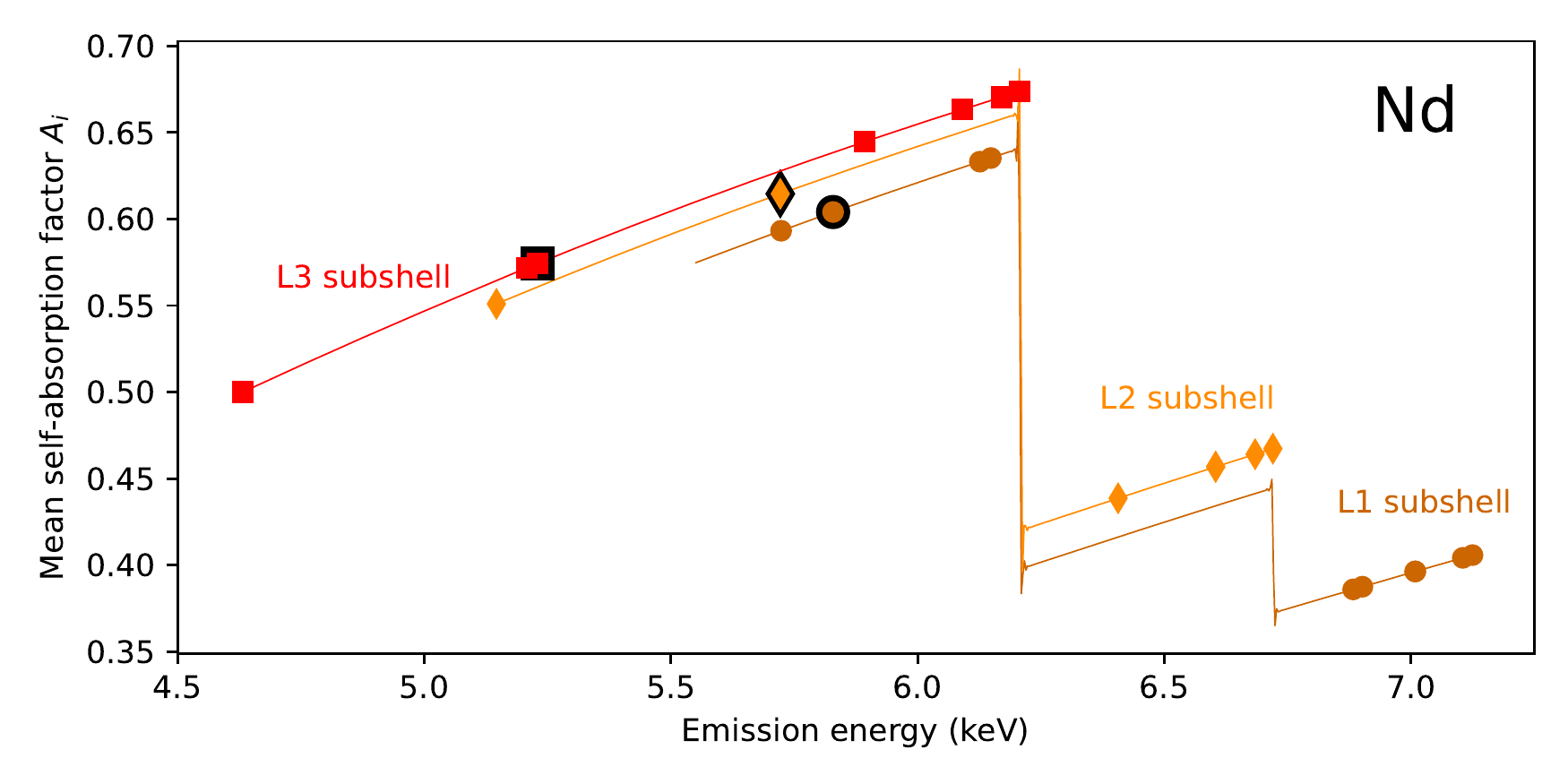}
    \caption{Self-absorption correction $A$ for L lines of Nd. Markers indicate the fluorescence lines detected from the L3 ($\blacksquare$), L2 ($\blacklozenge$), and  L1 ($\bullet$) subshells. The larger, black-outlined markers indicate the reference lines, the most intense line of each subshell. The curves show the form of the correction for all energies from the L$_x$M$_1$ line to the L$x$ edge. X~rays are absorbed more strongly above absorption edges, so a large drop in $A$ is seen for L2 lines at the L3 edge and for L1 lines at both the L2 and L3 edges (6.7\,keV and 6.2\,keV for Nd). The corrections for elements Pr, Tb, and Ho (not shown) are very similar.}
    \label{fig:sa_correction}
\end{figure}

The SA correction $A_i=A(E_i)$ can be found by numerical integration of Equation~\ref{eq:sa_weights}. The result will be a different function of fluorescence line energy $E_i$ for each element and for each subshell $x$. We use the CK factors from Campbell~\cite{Campbell1989} and the photoionization and total cross sections from {\tt xraylib}~\cite{Schoonjans2011} to compute it.  Figure~\ref{fig:sa_correction} shows the corrections for the three subshells of Nd. The fraction of emitted L-shell x~rays that escape the sample is between 0.37 and 0.70 for all the elements Pr, Nd, Tb, and Ho.

The measurement geometry is the largest source of systematic uncertainty on the self-absorption corrections and the only one in excess of 1\,\%. Our best estimate is that incoming and outgoing radiation make equal angles to the sample surface, $\theta=\phi=45^\circ$ (Figure~\ref{fig:geometry}). While the apparatus guarantees $\theta+\phi=90^\circ\pm2^\circ$, there is more uncertainty in the value of $\theta-\phi$. The freedom in the rotation of the sample holder and the mounting of samples on it means that we only know $\theta=45^\circ\pm20^\circ$. This geometrical uncertainty produces uncertainties on $A_i/A(E_\mathrm{ref})$ as high as 16\,\% on the L$_1$N, L$_1$O, L$_2$N and L$_2$O lines, and uncertainties of $\le7\,\%$ on the L3 and LM lines. Other uncertainties affect the relative $A$ by less than 0.5\,\%, including: the exact acceleration voltage in the x-ray tube; the ratio of bremsstrahlung to tungsten characteristic emission in the excitation spectrum; changes in the slope of $\widetilde{\mu}_x(E)$ by factors of $\pm1/3$; and changes in the slope of the total interaction cross section $\mu(E)$ by $\pm5\,\%$. We have also checked the stated uncertainties on the CK factors of Campbell~\cite{Campbell1989}. They change the relative $A$ by less than 0.1\,\%, as does replacement by the CK factors of Krause~\cite{Krause1979}.

\subsection{Summary of systematic uncertainties}

\begin{table}[tbh]
    \centering
    \begin{tabular}{lll}
          & This work & Optimized\\
         Cause & (worst case) & future system \\ \hline
         IR filter stack & 0.03 &  0.001 \\
         Air & 0.02 & 0.0001 \\
         $\mu$ for Au absorber & 0.01 & 0.0002 \\
         $\mu$ for Al (IR filter) & 0.008 & 0.0001 \\
         Absorber surface density & 0.002 & 0.0003 \\
         Absorber escape & 0.0003 & 0.0003 \\
         Vacuum window & 0.0001 & 0.0001 \\
         \hline
         Sample angle & 0.16 & 0.003 \\
         Tube voltage & 0.005 & 0.002 \\
         Tube W emission & 0.002 & 0.002 \\
         $\widetilde{\mu}(E)$ slope & 0.002 & 0.002 \\
         $\mu(E)$ slope $\pm5\,\%$  & 0.002 & 0.002 \\
         $S(E)$ slope & 0.001 & 0.001 \\
         CK factors & 0.0005 & 0.0005 \\
         Fluorescence yield & 0 & 0 \\ \hline
         Combined & 0.16 & 0.005 \\
    \end{tabular}
    \caption{Relative systematic uncertainties sorted by the size of the uncertainty achieved in the current measurement (\emph{This work}). The \emph{Optimized} column shows uncertainties from a hypothetical, optimized measurement with microcalorimeters outlined in Section~\ref{sec:prospects}. The first group of values are uncertainties in the spectrometer efficiency; the second group arises from the self-absorption correction. The specific values come from the most challenging line ratios measured here: Pr L$\ell$/L$\alpha_1$ for the DE model, and Nd L$\gamma_2$/L$\beta_3$ for the SA model. The last value (\emph{Combined}) is a quadrature sum of all other entries.}
    \label{tab:systematics}
\end{table}

Our main goal is to assess the uncertainties on the relative line intensities, both to understand what level has already been achieved and to project what uncertainties would be possible with a microcalorimeter measurement specifically dedicated to measurement of RFI values. Table~\ref{tab:systematics} summarizes this assessment. The DE uncertainty is as high as a few percent, while the self-absorption is up to 16\,\% uncertain. Section~\ref{sec:prospects} argues that a future measurement can reduce the combined uncertainty below 1\,\%.

\section{Results} \label{sec:results}

We have measured RFI values as part of a program to support quantitative modeling of x-ray emission spectra from fundamental atomic parameters. The L-line emission spectrum of an element can be considered the sum of three contributions, one from each L subshell. Depending on the excitation mechanism and spectrum, only L3 vacancies might be created, or only L3 and L2 vacancies. Even when vacancies are created in all three subshells, their proportions depend on the excitation source and on the absolute fluorescence yields, which are unequal for the three subshells and are not explored in this work.

It is assumed that regardless of how (for example) an L3 vacancy is produced, the L3 emission lines that result always appear in the same ratios. These ratios are the RFI sought in this work. For each subshell, we estimate the intensity of each fluorescence line relative to a reference line, which we choose to be the most intense line of the family:  L$\alpha_1$ (L$_3$M$_5$), L$\beta_1$ (L$_2$M$_4$), or L$\beta_3$ (L$_1$M$_3$).  The alternate approach, of estimating the branching fraction to each possible emission line out of a constrained total of 1, is also found in the literature. We prefer the reference-line approach because it limits the spreading of systematic uncertainties due to line ambiguities. For example, the L$\beta_3$ (an L1 line) and L$\beta_6$ (an L3 line) emission of Tb or Ho, which are not fully resolved, would cause correlated systematic uncertainties among all L1 and all L3 lines of these elements in a branching-fraction result. With ratios to reference-line intensities, however, uncertainties of this type are confined to the specific lines that are unresolved (provided the ambiguity does not involve the reference line).

Our RFI results appear in Table~\ref{tab:rfi_combined}. The standard uncertainties given in the table are the quadrature sum of statistical uncertainty on the photon counting, and systematics due to the uncertainty on the relative detection efficiency, the relative self-absorption correction, and ambiguities (if any) in allocating photons among unresolved lines. 

We emphasize that these results are derived from a pre-existing measurement that was optimized for the estimation of line energies and profiles rather than for line intensities. We show the complete table to demonstrate the potential power of microcalorimeters for RFI measurements, with at least 20 L lines analyzed per element, spanning a factor of 1000 in intensity.

Some so-called ``non-diagram lines'' identified in the spectra~\cite{Fowler2021} are included in Table~\ref{tab:rfi_combined}. They are identified with vacancies in a specific subshell by assumption that they are satellites of nearby lines of higher intensity. The non-diagram L$\gamma_{10}$ line is not resolved in the Tb and Ho spectra; for those samples, its intensity is included in the L$\gamma_2$ line.
 
\begin{table}[tb]
    \centering
    \begin{tabular}{llllll}
IUPAC & Siegbahn & Pr & Nd & Tb & Ho \\ \hline
    L$_3$M$_1$ & L$\ell$ & 0.0327(23) & 0.0339(24) & 0.0403(25) & 0.0431(26) \\
    L$_3$M$_4$ & L$\alpha_2$ & 0.1186(5) & 0.1143(4) & 0.1351(5) & 0.1265(5) \\
    L$_3$M$_5$ & L$\alpha_1$ & 1 & 1 & 1 & 1 \\
    L$_3$N$_1$ & L$\beta_6$ & 0.0125(26) & 0.0143(8) & 0.013(3) & 0.014(6) \\
    n/d  & L$\beta_{14}$ & 0.0110(24) & 0.0153(11) & 0.038(5) & 0.027(12) \\
    L$_3$N$_\mathrm{4,5}$ & L$\beta_{2,15}$ & 0.253(19) & 0.250(18) & 0.189(15) & 0.197(19) \\
    L$_3$O$_1$ & L$\beta_7$ & 0.0035(3) & 0.00193(19) & 0.0026(13) & 0.0039(3) \\
    L$_3$N$_\mathrm{6,7}$ & Lu & 0.00037(6) & 0.00108(11) & 0.00228(23) & 0.00058(6) \\
\hline
    L$_2$M$_1$ & L$\eta$ & 0.0212(26) & 0.028(4) & 0.0202(12) & 0.0250(13) \\
    L$_2$M$_4$ & L$\beta_1$ & 1 & 1 & 1 & 1 \\
    n/d  & L$\beta'$ &  & 0.0227(21) & 0.0212(22) & 0.0170(3) \\
    L$_2$N$_1$ & L$\gamma_5$ & 0.0072(12) & 0.0066(10) & 0.0070(9) & 0.0071(9) \\
    n/d  & L$\gamma_9$ & 0.038(7) & 0.041(18) & 0.074(8) & 0.088(24) \\
    L$_2$N$_4$ & L$\gamma_1$ & 0.174(28) & 0.158(28) & 0.135(15) & 0.121(25) \\
    L$_2$O$_1$ & L$\gamma_8$ & 0.0018(5) & 0.0020(3) & 0.00200(24) & 0.0025(3) \\
    L$_2$N$_\mathrm{6,7}$ & Lv & 0.00128(22) & 0.00158(24) & 0.00113(13) & 0.00103(13) \\
\hline
    L$_1$M$_2$ & L$\beta_4$ & 0.8(6) & 0.6(5) & 0.78(9) & 1.16(16) \\
    L$_1$M$_3$ & L$\beta_3$ & 1 & 1 & 1 & 1 \\
    L$_1$M$_4$ & L$\beta_{10}$ & 0.0084(10) & 0.0056(6) & 0.0168(20) & 0.023(3) \\
    L$_1$M$_5$ & L$\beta_{9}$ & 0.0381(17) & 0.0241(13) & 0.049(24) & 0.041(6) \\
    n/d  & L$\gamma_{10}$ & 0.18(3) & 0.17(4) &  &  \\
    L$_1$N$_2$ & L$\gamma_2$ & 0.120(23) & 0.08(3) & 0.33(27) & 0.23(7) \\
    L$_1$N$_3$ & L$\gamma_3$ & 0.19(4) & 0.17(4) & 0.19(3) & 0.22(6) \\
    L$_1$N$_\mathrm{4,5}$ & L$\gamma_{11}$ & 0.0056(9) & 0.0036(6) & 0.0045(9) & 0.0033(8) \\
    L$_1$O$_\mathrm{2,3}$ & L$\gamma_{4,4'}$ & 0.068(10) & 0.058(9) & 0.071(11) & 0.069(14) \\
    L$_1$N$_\mathrm{6,7}$ & - & 0.0057(15) & 0.0036(6) &  &  \\
\end{tabular}

    \caption{The fluorescence intensity for each detected emission line of Pr, Nd, Tb, and Ho. The IUPAC line name is given for diagram lines, or ``n/d'' for non-diagram lines. Intensities are given relative to the most intense line from each subshell. These reference lines are L$\alpha_1$, L$\beta_1$, and L$\beta_3$ respectively for subshells 3, 2, and 1. Intensities come from the measured number of photons, corrected for detector efficiency and for self-absorption in the lanthanide metal foil. Values in parentheses are the total uncertainty on the final digits, a quadrature sum of all systematic uncertainties and the statistical uncertainty on the photon counts. For several lines, an additional uncertainty is assigned due to the ambiguity in separating an unresolved neighbor (see online supplementary data for details including the separate components of uncertainty).
    }
    \label{tab:rfi_combined}
\end{table}

\subsection{Comparison to published results} \label{sec:comparison}

Unfortunately, there are few published measurements of L-line relative intensities for the rare-earth metals made with high-resolution spectrometers. We have three available sources of comparison: (1) ratios of line groups measured with lower-resolution spectrometers; (2) the interpolated theoretical values of Elam~\cite{Elam2002} and of Scofield~\cite{Scofield1974, Schoonjans2011}; and (3) two high-resolution diffractometer measurements. We attempt comparisons to verify that the RFI values of Table~\ref{tab:rfi_combined} are in the expected general range, and to establish the level of uncertainty found in the existing literature. The rich detail found in high-resolution spectra complicates these comparisons.

\begin{figure}[th]
    \centering
    \includegraphics[width=\textwidth]{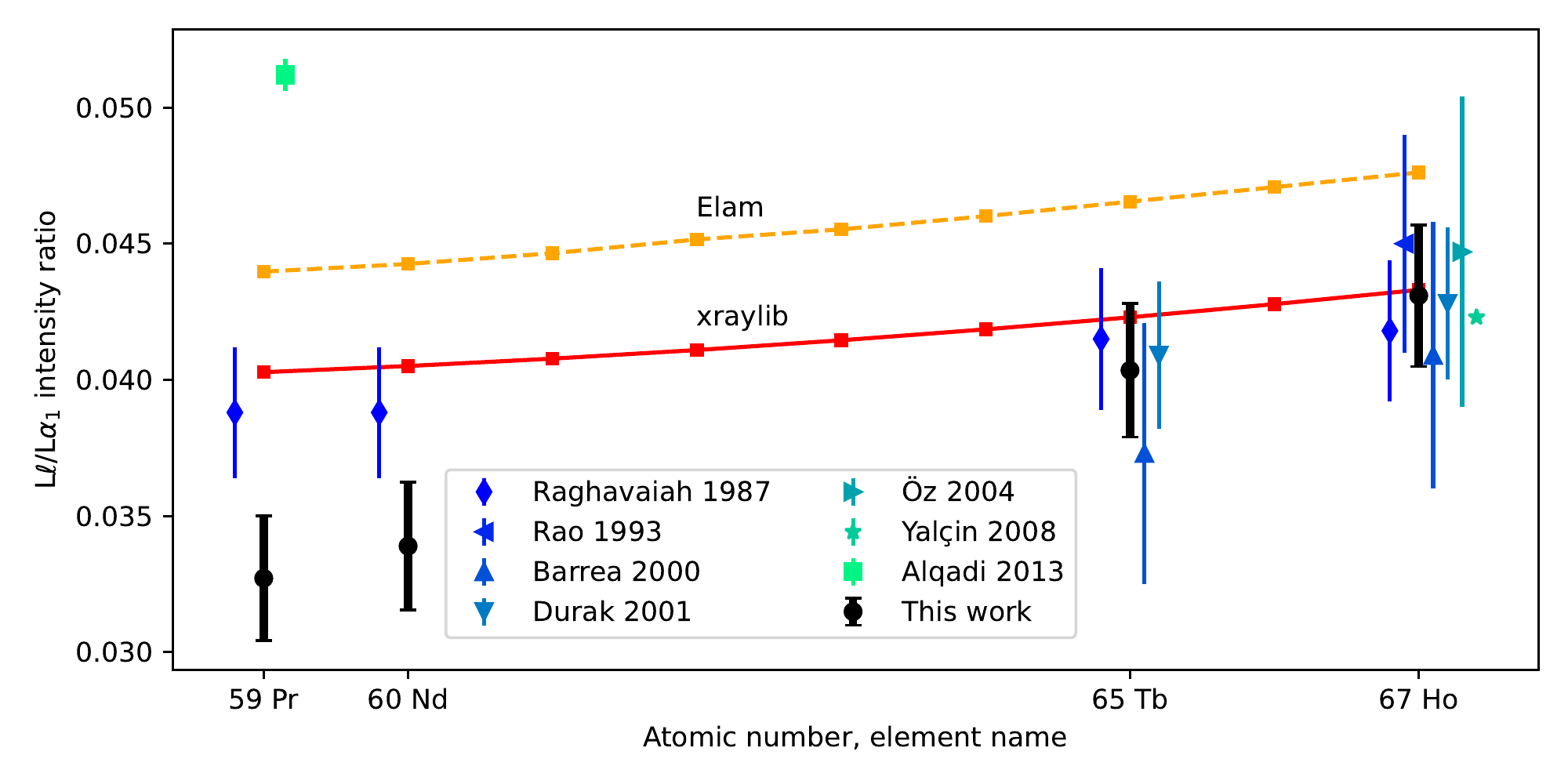}
    \caption{Published values of the L$\ell$/L$\alpha$ intensity ratio from two FP databases and several measurements. The values given for \emph{This work} use the L$\alpha_1$ intensity alone and would be approximately 10\,\% lower if the combined L$\alpha_{1,2}$ intensity was used for the denominator. Measurements are Raghavaiah~\cite{Raghavaiah1987}, Rao~\cite{Rao1993}, Barrea~\cite{Barrea2000}, Durak~\cite{Durak2001}, \"{O}z~\cite{Oz2004}, Yal\c{c}in~\cite{Yalcn2008}, and Alqadi~\cite{Alqadi2013}. Theory compilations are Elam~\cite{Elam2002} and xraylib~\cite{Schoonjans2011}.
    }
    \label{fig:Ll_ratios}
\end{figure}

Because the L$\gamma$ group includes L2 and L1 emission and the L$\beta$ group includes emission from all three subshells, we find no straightforward way to compare our data to published L$\gamma$ or L$\beta$ ratios. The L$\ell$/L$\alpha$ ratio, however, involves only L3 emission and permits direct comparison (Figure~\ref{fig:Ll_ratios}).  Still, the published experimental values all come from instruments with resolution of approximately 150\,eV, so the impact of the asymmetric L$\alpha_2$ feature on the L$\alpha$ intensity is hard to assess. Our Tb and Ho values are in agreement with prior measurements; the Pr and Nd values are somewhat lower.

Salem~\cite{Salem1971} and McClary~\cite{McCrary1972} have measured several line-intensity ratios that are relevant to our work. Direct comparisons are hard to make, however, given the unspecified wavelength resolution of these instruments and the authors' unspecified treatment of non-diagram lines and of overlapping lines. What we can say is that our 
L$\alpha_2$/L$\alpha_1$ and L$\beta_{2,15}$/L$\alpha_1$ ratios agree with Salem to the level of 10\,\% ($\sim 2$-$\sigma$), and our Ho L$\gamma_{4,4'}$/L$\beta_3$ and L$\gamma_{2,3}$/L$\beta_3$ ratios agree with McClary. Our estimates of L$\gamma_1$/L$\beta_1$ are somewhat lower than those given by either earlier work, but this is readily explained by our ability to resolve the L$\gamma_9$ non-diagram line from L$\gamma_1$ and our choice to treat it as a separate peak. Thus, in those few cases that admit a comparison to published measurements, our RFI estimates are broadly consistent with them. Further details appear in the online supplementary information.

\subsection{Prospects for future measurements} \label{sec:prospects}

Superconducting x-ray sensors with very high energy resolution offer great promise for RFI measurements. We explore here how a measurement similar to this one, but optimized for RFI, could perform. Table~\ref{tab:systematics} compares actual and projected uncertainties from the two cases.

A better DE estimate could be made with more careful characterization of the IR-blocking aluminum filters. Much thinner aluminum filters totaling 300\,nm thick could replace the 23\,$\mu$m trio used here, which would both increase the DE and reduce its uncertainty. Fully evacuating air from the sample chamber would also increase DE and improve the DE estimate. Thicker photon absorbers would also increase DE and reduce the effects of uncertain tabulated values of $\mu(E)$, with only a small penalty in the TESs' energy resolution. Values of DE around 80\,\% should be possible, which would improve the fluorescence detection rate by a factor of four relative to this data set, with a consequent reduction of statistical uncertainties by a factor of two for a fixed observation time and source flux. The higher efficiency would also be subject to much smaller uncertainties. The combination of tabulated mass-attenuation data with a benchmark measurement based on a source of known activity~\cite[for example]{Rodrigues2017} could also improve DE uncertainties. Future spectrometers are planned with reduced sensor-to-sensor crosstalk, which would also allow us to use all TESs in an array.

Uncertainty on the self-absorption correction is the limiting systematic on most of the relative intensities in Table~\ref{tab:rfi_combined}, so better self-absorption models would improve the results. One approach would be to use thin samples, less than 1\,$\mu$m thick, to reduce the size of the correction. Thin samples would cause a significant reduction in the fluorescence emission rates, though the high sensitivity and active area of TES arrays makes it possible to overcome this problem. Alternatively, we could use thick samples but with more careful control of the angular positioning (angles $\theta$ and $\phi$), though even here the potential roughness of the sample surfaces would remain a concern. The best approach might be to use both thin and thick samples. Comparison of the RFI for the more intense lines between samples would allow validation of the self-absorption model, while the thick sample would permit measurement of lines of lower intensity. In any scenario, we would control the sample orientation $\theta$ to $\pm2^\circ$.

Finally, it would be interesting to repeat this measurement with a range of photoexcitation sources. A monochromator could be used to simplify the SA correction, replacing the integral in Equation~\ref{eq:sa_weights} with a point estimate. A monochromatic source tuned below the L1 absorption edge of the sample would eliminate the L1 components from the spectrum, and an even lower energy (at the L2 edge) would also eliminate the L2 emission. Similar changes in the subshell ratio could be effected with a bremsstrahlung source by alteration of the tube voltage. Such data could reduce or eliminate the ambiguities arising from unresolved lines of different subshells, but a reduced voltage would also come at the cost of fluorescence intensity. In the other direction, an excitation source above 20\,keV, such as a gamma emitter or the K lines of cadmium or indium, would mix the three subshell spectral components, but it would do so in the high-energy limiting ratio that might be of widest interest. Electronic excitation, such as with a scanning electron microscope, could also be considered.

It is possible to measure different elements and energies with microcalorimeter spectrometers. The apparatus described in this work spans the approximate energy range of 4\,keV to 12\,keV, but the TES is not limited to this band. Sensors with thicker absorbers and higher heat capacity can extend the upper range above 100\,keV~\cite{Winkler2015}. In the other direction, smaller sensors with specialized vacuum window and IR-blocking filters can reach the 270\,eV carbon K line~\cite{Doriese:2017,Lee2019,Szypryt:2019}.

In addition to new measurements, we plan to perform computational modeling of the lanthanide elements with experts in modern atomic theory tools. A broad-band, high-resolution spectrum captured by a TES array offers an exciting opportunity to confront theory with highly constraining measurements.

\section{Conclusions}

The present study represents the first use of TES x-ray microcalorimeters to measure the relative intensities of L-series fluorescence lines, as far as we are aware. Mariam et al.~\cite{Mariam2022} have used other cryogenic microcalorimeters to perform a similar measurement of actinide L-lines at 11\,keV and above, excited by alpha and beta decays.  The strengths of this type of high-resolution cryogenic detector for RFI measurements include the ability to resolve even very close lines; a large collection area capable of measuring low-intensity lines; and simultaneous measurement across a band spanning at least a factor of two in energy. Instead of having to combine line groups and study the L$\beta$ family---combinations susceptible to excitation-dependent effects---we are able with the TES array to resolve and measure the intensities of specific transitions identified with well-defined subshell vacancies. The combination of these relative intensities with earlier results on line profiles makes a self-consistent data set, from which models of emission spectra can be made.

Improved control and measurement of experimental factors like the x-ray transmission of aluminum filters and the sample geometry, and the use of thinner sample films would all yield improved results. It should be possible by such steps to tame instrumental limitations and reach the level of 1\,\% systematic uncertainties, or better.

The ultimate limitation in an optimized measurement will be the discrimination of fluorescence lines from one another and from the background. The long Lorentzian tails and the unlimited possible asymmetries in atomic emission-line profiles make lines difficult to distinguish from background in a purely empirical fit. One can either accept an irreducible uncertainty on the line intensities, or try to use line shapes informed by theoretical calculations. 

Our long-term goal is the creation of a modern, complete, self-consistent, and SI-traceable database of x-ray fundamental parameters. With enough parameters, a fluorescence spectrum could be fully modeled. This goal would require parameters we have not attempted to measure with microcalorimeters, including photoionization cross sections, CK factors, and absolute fluorescence yields. Even without these, data on line profiles and line energies in combination with RFI values and a self-absorption correction are enough to model the fluorescence emission ``fingerprint'' of an element with only three unknown parameters: the absolute contribution of each subshell's spectrum. To fully realize this goal will likely require a diverse combination of measurement techniques, including wavelength-dispersive spectrometers and monochromatic x-ray sources with careful intensity calibrations~\cite{Honicke2014}.

With this study, we have shown that cryogenic microcalorimeters are capable of relative-intensity measurements across wide energy bands and over three orders of magnitude in line intensity. Future microcalorimeter measurements with realistic uncertainties of 1\,\% or better would be comparable in quality to almost all currently published values, while able to probe intensities from a far wider array of emission lines. Microcalorimeters can complement other techniques and will play an important role in the development of a 21st-century database of x-ray fundamental parameters.

{\vspace{2mm} \noindent \small
{\bf Author statement:} Joseph Fowler: Conceptualization, Formal analysis, Investigation, Writing. Luis Miaja-Avila: Investigation. Galen O'Neil: Conceptualization, Methodology, Investigation. Daniel Swetz: Conceptualization, Supervision, Project administration. Joel Ullom: Conceptualization, Supervision, Funding acquisition. Hope Whitelock: Software, Formal analysis.} 

{\vspace{2mm} \noindent \small
{\bf Declaration of competing interest:} The authors declare that they have no known competing financial interests or personal relationships that could have appeared to influence the work reported in this paper.}

{\vspace{2mm} \noindent \small
{\bf Acknowledgments:}
This work was supported by NIST's Innovations in Measurement Science program. We thank the European X-ray Spectrometry Association and the participants of its several FP Initiative workshops. We are grateful to 
Brad Alpert, 
Burkhard Beckhoff,
Randy Doriese, 
Mauro Guerra, 
Philipp H\"{o}nicke, 
Haibo Huang,
Larry Hudson, 
Young-Il Joe, 
Dan Schmidt, 
Csilla Szabo-Foster, and
Joel Weber
for helpful discussions and to three anonymous referees for helpful suggestions.  We thank the authors of {\tt xraylib} and of {\tt xraydb}, particularly Matt Newville, for their software packages that conveniently share x-ray fundamental parameters data.}

\bibliographystyle{elsarticle-num}
\bibliography{main}

\begin{thebibliography}{10}
\expandafter\ifx\csname url\endcsname\relax
  \def\url#1{\texttt{#1}}\fi
\expandafter\ifx\csname urlprefix\endcsname\relax\def\urlprefix{URL }\fi
\expandafter\ifx\csname href\endcsname\relax
  \def\href#1#2{#2} \def\path#1{#1}\fi

\bibitem{Salem1971}
S.~I. Salem, R.~T. Tsutsui, B.~A. Rabbani,
  \href{https://link.aps.org/doi/10.1103/PhysRevA.4.1728}{{L X-Ray Transition
  Probabilities in Elements with $Z\gtrsim 57$}}, Physical Review A 4~(5)
  (1971) 1728--1734.
\newblock

\bibitem{McCrary1972}
J.~H. McCrary, L.~V. Singman, L.~H. Ziegler, L.~D. Looney, C.~M. Edmonds, C.~E.
  Harris, \href{https://link.aps.org/doi/10.1103/PhysRevA.5.1587}{{L
  Fluorescent X-Ray Relative-Intensity Measurements}}, Physical Review A 5~(4)
  (1972) 1587--1591.
\newblock

\bibitem{Raghavaiah1987}
C.~V. Raghavaiah, N.~V. Rao, S.~B. Reddy, G.~Satyanarayana, D.~L. Sastry,
  \href{https://iopscience.iop.org/article/10.1088/0022-3700/20/21/015}{{L$\alpha$/L
  l X-ray intensity ratios for elements in the region 55$\le$Z$\le$80}},
  Journal of Physics B: Atomic and Molecular Physics 20~(21) (1987) 5647--5651.
\newblock

\bibitem{Raghavaiah1990}
C.~V. Raghavaiah, N.~V. Rao, S.~B. Reddy, G.~Satyanarayana, G.~S.~K. Murty,
  M.~V. S.~C. Rao, D.~L. Sastry,
  \href{https://onlinelibrary.wiley.com/doi/10.1002/xrs.1300190105}{{L$\alpha$/L$\beta$;
  and L$\alpha$/L$\gamma$ x-ray intensity ratios for elements in the range Z =
  55--80}}, X-Ray Spectrometry 19~(1) (1990) 23--26.
\newblock

\bibitem{Rao1993}
D.~V. Rao, R.~Cesareo, G.~E. Gigante,
  \href{https://link.aps.org/doi/10.1103/PhysRevA.47.1087}{{L x-ray
  fluorescence cross sections and intensity ratios in some high-Z elements
  excited by 23.62- and 24.68-keV photons}}, Physical Review A 47~(2) (1993)
  1087--1092.
\newblock

\bibitem{Barrea2000}
R.~A. Barrea, E.~V. Bonzi,
  \href{https://linkinghub.elsevier.com/retrieve/pii/S0969806X00003005}{{Experimental
  determination of L x-ray fluorescence cross-sections for rare earths at 10.70
  keV}}, Radiation Physics and Chemistry 59~(4) (2000) 347--354.
\newblock

\bibitem{Durak2001}
R.~Durak, Y.~{\"{O}}zdemir,
  \href{https://linkinghub.elsevier.com/retrieve/pii/S0375960101002237}{{Measurement
  of L$\alpha$/L$\ell$, L$\alpha$/L$\beta$ and L$\alpha$/L$\gamma$ X-ray
  intensity ratios for elements in the atomic range 57$\le$Z$\le$92 using
  radioisotope X-ray fluorescence}}, Physics Letters A 284~(1) (2001) 43--48.
\newblock

\bibitem{Oz2004}
E.~{\"{O}}z, E.~Bayda\c{s}, M.~Ertu\u{g}rul, Y.~\c{S}ahin,
  \href{http://link.springer.com/10.1023/B:JRNC.0000027064.64158.79}{{Measurement
  of L shell X-ray fluorescence intensity ratios for some elements in the
  atomic number range of 66{$\le$}Z{$\le$}90 by photoionization of consecutive
  L-subshells}}, Journal of Radioanalytical and Nuclear Chemistry 260~(1)
  (2004) 75--79.
\newblock

\bibitem{Salah2005}
W.~Salah, J.~Al-Jundi,
  \href{https://linkinghub.elsevier.com/retrieve/pii/S0022407304004133}{{Measurement
  of L X-ray cross-sections and relative intensities of heavy elements by
  15.2\,keV photons}}, Journal of Quantitative Spectroscopy and Radiative
  Transfer 94~(3-4) (2005) 325--333.
\newblock

\bibitem{Yalcn2008}
P.~Yal{\c{c}}ın, S.~Porikli, Y.~Kurucu, Y.~{\c{S}}ahin,
  \href{https://linkinghub.elsevier.com/retrieve/pii/S0370269308004620}{{Measurement
  of relative L X-ray intensity ratio following radioactive decay and
  photoionization}}, Physics Letters B 663~(3) (2008) 186--190.
\newblock

\bibitem{Alqadi2013}
M.~Alqadi, Y.~Alsenjlawi, F.~Alzoubi,
  \href{https://linkinghub.elsevier.com/retrieve/pii/S0969806X13001011}{{Measurement
  of L X-Ray relative intensities for selected heavy elements}}, Radiation
  Physics and Chemistry 87 (2013) 31--34.
\newblock

\bibitem{Garg1984}
M.~L. Garg, J.~Singh, H.~R. Verma, N.~Singh, P.~C. Mangal, P.~N. Trehan,
  \href{https://iopscience.iop.org/article/10.1088/0022-3700/17/4/013}{{Relative
  intensity measurements of L-shell X-rays for Ta, Au, Pb and Bi in the energy
  range 17-60 keV}}, Journal of Physics B: Atomic and Molecular Physics 17~(4)
  (1984) 577--584.
\newblock

\bibitem{Rao1995}
D.~Rao, R.~Cesareo, G.~Gigante,
  \href{https://linkinghub.elsevier.com/retrieve/pii/0969806X9400113X}{{L X-ray
  fluorescence cross sections, fluorescence yields and intensity ratios for Au
  and Pb at excitation energies 21.56, 31.64 and 34.17 keV}}, Radiation Physics
  and Chemistry 46~(1) (1995) 17--22.
\newblock

\bibitem{Ismail2000}
A.~M. Ismail, N.~B. Malhi,
  \href{https://onlinelibrary.wiley.com/doi/10.1002/1097-4539(200007/08)29:4%3C317::AID-XRS437%3E3.0.CO;2-C}{{L-shell
  x-ray relative intensities of some heavy elements excited by 20.48 keV
  x-rays}}, X-Ray Spectrometry 29~(4) (2000) 317--319.
\newblock

\bibitem{Gurol2003}
A.~G{\"{u}}rol, A.~Karabulut,
  \href{https://linkinghub.elsevier.com/retrieve/pii/S0584854703000892}{{L
  subshell fluorescence cross-sections and relative intensity ratios of some
  elements in the atomic range 72$\le$Z$\le$92}}, Spectrochimica Acta Part B:
  Atomic Spectroscopy 58~(8) (2003) 1473--1480.
\newblock

\bibitem{Demir2008}
L.~Demir, I.~Han, M.~Şahin,
  \href{https://linkinghub.elsevier.com/retrieve/pii/S0368204807002733}{{Measurement
  of L X-ray fluorescence cross sections and relative intensities for some
  elements in the atomic range 78$\le$Z$\le$92}}, Journal of Electron
  Spectroscopy and Related Phenomena 162~(1) (2008) 44--48.
\newblock

\bibitem{Durdu2012}
B.~Durdu, A.~Kucukonder,
  \href{https://linkinghub.elsevier.com/retrieve/pii/S0969806X11003641}{{Variation
  of the L X-ray fluorescence cross-sections, intensity ratios and fluorescence
  yields of Sm and Eu in halogen compounds}}, Radiation Physics and Chemistry
  81~(2) (2012) 135--142.
\newblock

\bibitem{Wang2015}
X.~Wang, Z.~Xu, L.~Zhang,
  \href{https://linkinghub.elsevier.com/retrieve/pii/S0969806X15001206}{{L
  X-ray intensity ratios for high Z elements induced with X-ray tube}},
  Radiation Physics and Chemistry 112 (2015) 121--124.
\newblock

\bibitem{Ganly2016}
B.~Ganly, Y.~{Van Haarlem}, J.~Tickner,
  \href{https://onlinelibrary.wiley.com/doi/abs/10.1002/xrs.2695}{{Measurement
  of relative line intensities for L‐shell X‐rays from selected elements
  between Z= 68 (Er) and Z= 79 (Au)}}, X-ray Spectrometry 45~(4) (2016)
  233--243.
\newblock

\bibitem{Durdag2017}
S.~P. Durdağı,
  \href{https://linkinghub.elsevier.com/retrieve/pii/S0026265X16302399}{{Chemical
  environment change analysis on L X-ray emission spectra of some lanthanide
  compounds}}, Microchemical Journal 130 (2017) 27--32.
\newblock

\bibitem{Menesguen2018}
Y.~M{\'{e}}nesguen, M.-C. L{\'{e}}py, J.~M. Sampaio, J.~P. Marques, F.~Parente,
  M.~Guerra, P.~Indelicato, J.~P. Santos,
  \href{https://iopscience.iop.org/article/10.1088/1681-7575/aad1d6}{{Experimental
  and theoretical determination of the L-fluorescence yields of bismuth}},
  Metrologia 55~(5) (2018) 621--630.
\newblock

\bibitem{Wansleben2019}
M.~Wansleben, Y.~Kayser, P.~H{\"{o}}nicke, I.~Holfelder, A.~W{\"{a}}hlisch,
  R.~Unterumsberger, B.~Beckhoff,
  \href{https://iopscience.iop.org/article/10.1088/1681-7575/ab40d2}{{Experimental
  determination of line energies, line widths and relative transition
  probabilities of the Gadolinium L x-ray emission spectrum}}, Metrologia
  56~(6) (2019) 065007.
\newblock

\bibitem{Mariam2022}
R.~Mariam, M.~Rodrigues, M.~Loidl, S.~Pierre, V.~Lourenço,
  \href{https://doi.org/10.1016/j.sab.2021.106331}{{Determination of L-X ray
  absolute emission intensities of 238Pu, 244Cm, 237Np and 233Pa radionuclides
  using a metallic magnetic calorimeter}}, Spectrochimica Acta Part B: Atomic
  Spectroscopy 187 (2022) 106331.
\newblock

\bibitem{Salem1974}
S.~Salem, S.~Panossian, R.~Krause,
  \href{https://linkinghub.elsevier.com/retrieve/pii/S0092640X74800173}{{Experimental
  K and L relative x-ray emission rates}}, Atomic Data and Nuclear Data Tables
  14~(2) (1974) 91--109.
\newblock

\bibitem{Elam2002}
W.~Elam, B.~Ravel, J.~Sieber,
  \href{https://www.sciencedirect.com/science/article/pii/S0969806X01002274?via%3Dihub}{{A
  new atomic database for X-ray spectroscopic calculations}}, Radiation Physics
  and Chemistry 63~(2) (2002) 121--128.
\newblock

\bibitem{Scofield1974}
J.~H. Scofield,
  \href{https://linkinghub.elsevier.com/retrieve/pii/S0092640X74800197}{{Relativistic
  Hartree-Slater values for K and L X-ray emission rates}}, Atomic Data and
  Nuclear Data Tables 14~(2) (1974) 121--137.
\newblock

\bibitem{Schoonjans2011}
T.~Schoonjans, A.~Brunetti, B.~Golosio, M.~{Sanchez del Rio}, V.~A. Sol{\'{e}},
  C.~Ferrero, L.~Vincze,
  \href{https://linkinghub.elsevier.com/retrieve/pii/S0584854711001984}{{The
  xraylib library for X-ray–matter interactions. Recent developments}},
  Spectrochimica Acta Part B: Atomic Spectroscopy 66~(11-12) (2011) 776--784.
\newblock

\bibitem{Fowler2021}
J.~W. Fowler, G.~C. O'Neil, B.~K. Alpert, D.~A. Bennett, E.~V. Denison, W.~B.
  Doriese, G.~C. Hilton, L.~T. Hudson, Y.-I. Joe, K.~M. Morgan, D.~R. Schmidt,
  D.~S. Swetz, C.~I. Szabo, J.~N. Ullom,
  \href{https://iopscience.iop.org/article/10.1088/1681-7575/abd28a}{{Absolute
  energies and emission line shapes of the L x-ray transitions of lanthanide
  metals}}, Metrologia 58~(1) (2021) 015016.
\newblock

\bibitem{Ullom2015}
J.~N. Ullom, D.~A. Bennett,
  \href{http://iopscience.iop.org/article/10.1088/0953-2048/28/8/084003}{{Review
  of superconducting transition-edge sensors for x-ray and gamma-ray
  spectroscopy}}, Superconductor Science and Technology 28~(8) (2015) 84003.
\newblock

\bibitem{Doriese:2017}
W.~Doriese, P.~Abbamonte, B.~K. Alpert, D.~Bennett, E.~Denison, Y.~Fang,
  D.~Fischer, C.~Fitzgerald, J.~W. Fowler, J.~Gard, J.~Hays-Wehle, G.~Hilton,
  C.~Jaye, J.~McChesney, L.~Miaja-Avila, K.~Morgan, Y.~Joe, G.~O'Neil,
  C.~Reintsema, F.~Rodolakis, D.~Schmidt, H.~Tatsuno, J.~Uhlig, L.~Vale,
  J.~Ullom, D.~Swetz, {A practical superconducting-microcalorimeter X-ray
  spectrometer for beamline and laboratory science}, Review of Scientific
  Instruments 88~(5) (2017).
\newblock

\bibitem{Doriese:2016}
W.~Doriese, K.~Morgan, D.~Bennett, E.~Denison, C.~Fitzgerald, J.~W. Fowler,
  J.~Gard, J.~Hays-Wehle, G.~Hilton, K.~Irwin, Y.~Joe, J.~Mates, G.~O'Neil,
  C.~Reintsema, N.~Robbins, D.~Schmidt, D.~Swetz, H.~Tatsuno, L.~Vale,
  J.~Ullom,
  \href{https://link.springer.com/article/10.1007%2Fs10909-015-1373-z}{{Developments
  in Time-Division Multiplexing of X-ray Transition-Edge Sensors}}, Journal of
  Low Temperature Physics 184~(1-2) (2016) 389--395.
\newblock

\bibitem{Holzer1997}
G.~H{\"{o}}lzer, M.~Fritsch, M.~Deutsch, J.~H{\"{a}}rtwig, E.~F{\"{o}}rster,
  \href{http://pra.aps.org/abstract/PRA/v56/i6/p4554_1}{{K$\alpha$1,2 and
  K$\beta$1,3 x-ray emission lines of the 3d transition metals}}, Physical
  Review A 56~(6) (1997) 4554--4568.
\newblock

\bibitem{Mendenhall2017}
M.~H. Mendenhall, A.~Henins, L.~T. Hudson, C.~I. Szabo, D.~Windover, J.~P.
  Cline,
  \href{http://stacks.iop.org/0953-4075/50/i=11/a=115004?key=crossref.f53c39b7d2490e21ab1259f4078c6d65}{{High-precision
  measurement of the x-ray Cu K$\alpha$ spectrum}}, Journal of Physics B:
  Atomic, Molecular and Optical Physics 50~(11) (2017) 115004.
\newblock

\bibitem{Chantler2006}
C.~T. Chantler, M.~N. Kinnane, C.-H. Su, J.~A. Kimpton,
  \href{https://link.aps.org/doi/10.1103/PhysRevA.73.012508}{Characterization
  of k$\alpha$ spectral profiles for vanadium, component redetermination for
  scandium, titanium, chromium, and manganese, and development of satellite
  structure for {$Z=21$} to {$Z=25$}}, Physical Review A 73 (2006) 012508.
\newblock

\bibitem{Chantler2013}
C.~T. Chantler, J.~A. Lowe, I.~P. Grant,
  \href{https://iopscience.iop.org/article/10.1088/0953-4075/46/1/015002}{{High-accuracy
  reconstruction of titanium x-ray emission spectra, including relative
  intensities, asymmetry and satellites, and ab initio determination of shake
  magnitudes for transition metals}}, Journal of Physics B: Atomic, Molecular
  and Optical Physics 46~(1) (2013) 015002.
\newblock

\bibitem{Fowler2022}
J.~W. Fowler, B.~K. Alpert, G.~C. O'Neil, D.~S. Swetz, J.~N. Ullom,
  \href{http://arxiv.org/abs/2204.08431}{{Energy calibration of nonlinear
  microcalorimeters with uncertainty estimates from Gaussian process
  regression}}, Journal of Low Temperature Physics (apr 2022).
\newblock \href {http://arxiv.org/abs/2204.08431} {\path{arXiv:2204.08431}},

\bibitem{Lovell1968}
S.~Lovell, E.~Rollinson,
  \href{https://www.nature.com/articles/2181179a0}{{Density of Thin Films of
  Vacuum Evaporated Metals}}, Nature 218~(5147) (1968) 1179--1180.
\newblock

\bibitem{Siegel2011}
J.~Siegel, O.~Lyutakov, V.~Rybka, Z.~Kolsk{\'{a}},
  V.~{\v{S}}vor{\v{c}}{\'{i}}k,
  \href{https://nanoscalereslett.springeropen.com/articles/10.1186/1556-276X-6-96}{{Properties
  of gold nanostructures sputtered on glass}}, Nanoscale Research Letters 6~(1)
  (2011) 96.
\newblock

\bibitem{Coster1935}
D.~Coster, R.~D. {L. Kronig},
  \href{https://linkinghub.elsevier.com/retrieve/pii/S003189143590060X}{{New
  type of auger effect and its influence on the x-ray spectrum}}, Physica
  2~(1-12) (1935) 13--24.
\newblock

\bibitem{Campbell1989}
J.~Campbell, J.-X. Wang,
  \href{https://linkinghub.elsevier.com/retrieve/pii/0092640X89900041}{{Interpolated
  Dirac-Fock values of L-subshell x-ray emission rates including overlap and
  exchange effects}}, Atomic Data and Nuclear Data Tables 43~(2) (1989)
  281--291.
\newblock

\bibitem{Krause1979}
M.~O. Krause, \href{http://aip.scitation.org/doi/10.1063/1.555594}{{Atomic
  radiative and radiationless yields for K and L shells}}, Journal of Physical
  and Chemical Reference Data 8~(2) (1979) 307--327.
\newblock

\bibitem{Rodrigues2017}
M.~Rodrigues, R.~Mariam, M.~Loidl,
  \href{http://www.epj-conferences.org/10.1051/epjconf/201714610012}{{A
  metallic magnetic calorimeter dedicated to the spectrometry of L X-rays
  emitted by actinides}}, EPJ Web of Conferences 146 (2017) 10012.
\newblock

\bibitem{Winkler2015}
R.~Winkler, A.~Hoover, M.~Rabin, D.~Bennett, W.~Doriese, J.~Fowler,
  J.~Hays-Wehle, R.~Horansky, C.~Reintsema, D.~Schmidt, L.~Vale, J.~Ullom,
  \href{https://linkinghub.elsevier.com/retrieve/pii/S0168900214010778}{256-pixel
  microcalorimeter array for high-resolution $\gamma$-ray spectroscopy of
  mixed-actinide materials}, Nuclear Instruments and Methods in Physics
  Research Section A: Accelerators, Spectrometers, Detectors and Associated
  Equipment 770 (2015) 203--210.
\newblock

\bibitem{Lee2019}
S.-J. Lee, C.~J. Titus, R.~{Alonso Mori}, M.~L. Baker, D.~A. Bennett, H.-M.
  Cho, W.~B. Doriese, J.~W. Fowler, K.~J. Gaffney, A.~Gallo, J.~D. Gard, G.~C.
  Hilton, H.~Jang, Y.~I. Joe, C.~J. Kenney, J.~Knight, T.~Kroll, J.-S. Lee,
  D.~Li, D.~Lu, R.~Marks, M.~P. Minitti, K.~M. Morgan, H.~Ogasawara, G.~C.
  O'Neil, C.~D. Reintsema, D.~R. Schmidt, D.~Sokaras, J.~N. Ullom, T.-C. Weng,
  C.~Williams, B.~A. Young, D.~S. Swetz, K.~D. Irwin, D.~Nordlund,
  \href{http://aip.scitation.org/doi/10.1063/1.5119155}{{Soft X-ray
  spectroscopy with transition-edge sensors at Stanford Synchrotron Radiation
  Lightsource beamline 10-1}}, Review of Scientific Instruments 90~(11) (2019)
  113101.
\newblock

\bibitem{Szypryt:2019}
P.~Szypryt, G.~C. O'Neil, E.~Takacs, J.~N. Tan, S.~W. Buechele, A.~S. Naing,
  D.~A. Bennett, W.~B. Doriese, M.~Durkin, J.~W. Fowler, J.~D. Gard, G.~C.
  Hilton, K.~M. Morgan, C.~D. Reintsema, D.~R. Schmidt, D.~S. Swetz, J.~N.
  Ullom, Y.~Ralchenko, \href{http://aip.scitation.org/doi/10.1063/1.5116717}{{A
  transition-edge sensor-based x-ray spectrometer for the study of highly
  charged ions at the National Institute of Standards and Technology electron
  beam ion trap}}, Review of Scientific Instruments 90~(12) (2019) 123107.
\newblock

\bibitem{Honicke2014}
P.~H{\"{o}}nicke, M.~Kolbe, M.~M{\"{u}}ller, M.~Mantler, M.~Kr{\"{a}}mer,
  B.~Beckhoff,
  \href{https://link.aps.org/doi/10.1103/PhysRevLett.113.163001}{{Experimental
  Verification of the Individual Energy Dependencies of the Partial L-Shell
  Photoionization Cross Sections of Pd and Mo}}, Physical Review Letters
  113~(16) (2014) 163001.
\newblock

\end{thebibliography}

\newpage
\appendix
\section{Online Supplement}

\renewcommand{\thetable}{\Alph{table}}
\setcounter{table}{0}

\alert{This section and its five tables (\ref{tab:results_Pr} to \ref{tab:salem_compare}) should be offered as an online-only supplement.}

\begin{table}[tb]
    \centering
    \begin{tabular}{llrrccccccc}
\multicolumn{2}{c}{Pr emission line} & Energy & Photons & & & Stat. &  \multicolumn{3}{c}{\dotfill Systematic \dotfill} & Comb. \\
IUPAC & Sieg. & (eV) & counted & DE & SA & uncert. & $\delta$DE & $\delta$SA & $\delta$LS & uncert.  \\ \hline
    L$_3$M$_1$ & L$\ell$ & 4457.6 & 39300 & 0.101 & 0.495 & 0.006 & 0.032 & 0.055 & 0.03 & 0.070 \\
    L$_3$M$_4$ & L$\alpha_2$ & 5011.9 & 242200 & 0.150 & 0.566 & 0.004 & 0.001 & 0.002 &  & 0.004 \\
    L$_3$M$_5$ & L$\alpha_1$ & 5032.9 & 2072000 & 0.152 & 0.568 & 0.001 & --- & --- &  & 0.001 \\
    L$_3$N$_1$ & L$\beta_6$ & 5660.2 & 35900 & 0.188 & 0.638 & 0.012 & 0.022 & 0.054 & 0.20 & 0.209 \\
    n/d  & L$\beta_{14}$ & 5830.1 & 33400 & 0.194 & 0.654 & 0.032 & 0.026 & 0.067 & 0.20 & 0.215 \\
    L$_3$N$_\mathrm{4,5}$ & L$\beta_{2,15}$ & 5849.7 & 777000 & 0.194 & 0.656 & 0.002 & 0.027 & 0.068 & 0.01 & 0.074 \\
    L$_3$O$_1$ & L$\beta_7$ & 5925.1 & 11000 & 0.197 & 0.663 & 0.043 & 0.029 & 0.074 &  & 0.090 \\
    L$_3$N$_\mathrm{6,7}$ & Lu & 5960.6 & 1200 & 0.198 & 0.667 & 0.138 & 0.030 & 0.077 &  & 0.160 \\
\hline
    L$_2$M$_1$ & L$\eta$ & 4933.4 & 24600 & 0.144 & 0.542 & 0.021 & 0.021 & 0.048 &  & 0.057 \\
    L$_2$M$_4$ & L$\beta_1$ & 5488.5 & 1621000 & 0.180 & 0.606 & 0.002 & --- & --- & 0.11 & 0.110 \\
    L$_2$N$_1$ & L$\gamma_5$ & 6136.2 & 9200 & 0.201 & 0.428 & 0.017 & 0.016 & 0.131 &  & 0.133 \\
    n/d  & L$\gamma_9$ & 6305.4 & 51100 & 0.203 & 0.444 & 0.008 & 0.020 & 0.119 & 0.08 & 0.145 \\
    L$_2$N$_4$ & L$\gamma_1$ & 6322.2 & 233200 & 0.203 & 0.445 & 0.003 & 0.020 & 0.118 & 0.02 & 0.122 \\
    L$_2$O$_1$ & L$\gamma_8$ & 6402.0 & 2400 & 0.204 & 0.453 & 0.067 & 0.021 & 0.113 & 0.20 & 0.240 \\
    L$_2$N$_\mathrm{6,7}$ & Lv & 6437.5 & 1770 & 0.204 & 0.456 & 0.051 & 0.022 & 0.111 & 0.03 & 0.127 \\
\hline
    L$_1$M$_2$ & L$\beta_4$ & 5496.8 & 220000 & 0.181 & 0.582 & 0.014 & 0.003 & 0.008 & 0.80 & 0.800 \\
    L$_1$M$_3$ & L$\beta_3$ & 5593.0 & 295000 & 0.185 & 0.592 & 0.003 & --- & --- & 0.01 & 0.013 \\
    L$_1$M$_4$ & L$\beta_{10}$ & 5883.5 & 2700 & 0.195 & 0.622 & 0.112 & 0.008 & 0.023 &  & 0.115 \\
    L$_1$M$_5$ & L$\beta_{9}$ & 5902.7 & 12500 & 0.196 & 0.624 & 0.033 & 0.008 & 0.024 &  & 0.042 \\
    n/d  & L$\gamma_{10}$ & 6579.3 & 36300 & 0.204 & 0.372 & 0.011 & 0.021 & 0.159 & 0.10 & 0.189 \\
    L$_1$N$_2$ & L$\gamma_2$ & 6601.0 & 24600 & 0.204 & 0.373 & 0.014 & 0.022 & 0.158 & 0.10 & 0.188 \\
    L$_1$N$_3$ & L$\gamma_3$ & 6617.1 & 39600 & 0.204 & 0.375 & 0.007 & 0.022 & 0.157 & 0.10 & 0.187 \\
    L$_1$N$_\mathrm{4,5}$ & L$\gamma_{11}$ & 6718.3 & 1170 & 0.204 & 0.383 & 0.055 & 0.023 & 0.151 &  & 0.162 \\
    L$_1$O$_\mathrm{2,3}$ & L$\gamma_{4,4'}$ & 6814.0 & 14500 & 0.203 & 0.391 & 0.011 & 0.025 & 0.145 & 0.02 & 0.149 \\
    L$_1$N$_\mathrm{6,7}$ & - & 6829.2 & 1210 & 0.203 & 0.393 & 0.071 & 0.025 & 0.144 & 0.20 & 0.258 \\
\end{tabular}

    \caption{Fluorescent-intensity measurements for the Pr foil sample, 250\,$\mu$m thick, 99.5\,\% pure (with 0.4\,\% Nd). \emph{Energy} gives the emission line energy previously published~\cite{Fowler2021}. \emph{Photons counted} is the total number of photons detected in an emission line. It was estimated by fitting the observed emission spectrum. \emph{DE} and \emph{SA} are the detection and self-absorption efficiencies, respectively. The RFI that appears in Table~\ref{tab:rfi_combined} equals the photons counted divided by (DE$\times$SA), relative to a reference line. \emph{Stat.\ uncert.\ }is the fractional uncertainty on the \emph{Photons counted} estimate.  The three columns \emph{$\delta$DE}, \emph{$\delta$SA}, \emph{$\delta$LS} are the fractional uncertainties on the relative detection efficiency, the relative SA efficiency, and line separation ambiguities. The first two are given as ``---'' for the reference lines in each subshell. The \emph{Comb.\ uncert.} is the total fractional uncertainty, the quadrature combination of the three systematic terms and the (fractional) statistical uncertainty on the measured intensity.  The three sets of rows in the table correspond to emission from the L3, L2, and L1 subshells.
    }
    \label{tab:results_Pr}
\end{table}

\begin{table}[tb]
    \centering
    \begin{tabular}{llrrccccccc}
\multicolumn{2}{c}{Nd emission line} & Energy & Photons & & & Stat. &  \multicolumn{3}{c}{\dotfill Systematic \dotfill} & Comb. \\
IUPAC & Sieg. & (eV) & counted & DE & SA & uncert. & $\delta$DE & $\delta$SA & $\delta$LS & uncert.  \\ \hline
    L$_3$M$_1$ & L$\ell$ & 4632.0 & 45100 & 0.118 & 0.500 & 0.006 & 0.028 & 0.055 & 0.03 & 0.069 \\
    L$_3$M$_4$ & L$\alpha_2$ & 5207.3 & 242300 & 0.164 & 0.572 & 0.003 & 0.001 & 0.002 &  & 0.004 \\
    L$_3$M$_5$ & L$\alpha_1$ & 5229.9 & 2148000 & 0.166 & 0.574 & 0.001 & --- & --- &  & 0.001 \\
    L$_3$N$_1$ & L$\beta_6$ & 5891.6 & 40600 & 0.196 & 0.645 & 0.008 & 0.020 & 0.055 &  & 0.059 \\
    n/d  & L$\beta_{14}$ & 6069.3 & 45700 & 0.200 & 0.661 & 0.019 & 0.024 & 0.068 &  & 0.075 \\
    L$_3$N$_\mathrm{4,5}$ & L$\beta_{2,15}$ & 6090.4 & 749000 & 0.200 & 0.663 & 0.002 & 0.024 & 0.070 &  & 0.074 \\
    L$_3$O$_1$ & L$\beta_7$ & 6170.2 & 5900 & 0.201 & 0.670 & 0.062 & 0.026 & 0.076 &  & 0.101 \\
    L$_3$N$_\mathrm{6,7}$ & Lu & 6206.8 & 3300 & 0.202 & 0.674 & 0.066 & 0.027 & 0.078 &  & 0.106 \\
\hline
    L$_2$M$_1$ & L$\eta$ & 5145.0 & 33600 & 0.160 & 0.551 & 0.020 & 0.019 & 0.049 & 0.10 & 0.115 \\
    n/d  & L$\beta'$ & 5737.5 & 36600 & 0.191 & 0.616 & 0.014 & 0.000 & 0.001 &  & 0.015 \\
    L$_2$M$_4$ & L$\beta_1$ & 5721.0 & 1601500 & 0.190 & 0.614 & 0.001 & --- & --- & 0.09 & 0.090 \\
    L$_2$N$_1$ & L$\gamma_5$ & 6405.0 & 8020 & 0.204 & 0.439 & 0.017 & 0.015 & 0.130 &  & 0.132 \\
    n/d  & L$\gamma_9$ & 6580.7 & 52600 & 0.204 & 0.455 & 0.008 & 0.018 & 0.118 & 0.40 & 0.418 \\
    L$_2$N$_4$ & L$\gamma_1$ & 6600.4 & 202200 & 0.204 & 0.457 & 0.003 & 0.018 & 0.117 & 0.10 & 0.155 \\
    L$_2$O$_1$ & L$\gamma_8$ & 6682.5 & 2610 & 0.204 & 0.464 & 0.047 & 0.019 & 0.112 &  & 0.123 \\
    L$_2$N$_\mathrm{6,7}$ & Lv & 6721.3 & 2060 & 0.204 & 0.467 & 0.041 & 0.020 & 0.109 &  & 0.119 \\
\hline
    L$_1$M$_2$ & L$\beta_4$ & 5721.0 & 177900 & 0.190 & 0.593 & 0.001 & 0.003 & 0.008 & 0.81 & 0.810 \\
    L$_1$M$_3$ & L$\beta_3$ & 5827.9 & 306200 & 0.194 & 0.604 & 0.003 & --- & --- & 0.01 & 0.010 \\
    L$_1$M$_4$ & L$\beta_{10}$ & 6125.4 & 1900 & 0.201 & 0.633 & 0.096 & 0.007 & 0.023 &  & 0.099 \\
    L$_1$M$_5$ & L$\beta_{9}$ & 6147.0 & 8000 & 0.201 & 0.635 & 0.045 & 0.007 & 0.024 &  & 0.052 \\
    n/d  & L$\gamma_{10}$ & 6862.9 & 34500 & 0.203 & 0.384 & 0.011 & 0.019 & 0.159 & 0.15 & 0.220 \\
    L$_1$N$_2$ & L$\gamma_2$ & 6882.8 & 15500 & 0.203 & 0.386 & 0.019 & 0.019 & 0.158 & 0.40 & 0.431 \\
    L$_1$N$_3$ & L$\gamma_3$ & 6899.9 & 34300 & 0.202 & 0.387 & 0.007 & 0.019 & 0.157 & 0.15 & 0.218 \\
    L$_1$N$_\mathrm{4,5}$ & L$\gamma_{11}$ & 7007.1 & 740 & 0.201 & 0.396 & 0.068 & 0.021 & 0.151 &  & 0.167 \\
    L$_1$O$_\mathrm{2,3}$ & L$\gamma_{4,4'}$ & 7106.9 & 12400 & 0.200 & 0.404 & 0.012 & 0.022 & 0.145 &  & 0.147 \\
    L$_1$N$_\mathrm{6,7}$ & - & 7121.4 & 760 & 0.200 & 0.406 & 0.097 & 0.022 & 0.144 &  & 0.175 \\
\end{tabular}

    \caption{Fluorescent-intensity measurements for the Nd foil sample, 300\,$\mu$m thick, 99.9\,\% pure. Columns are as in Table~\ref{tab:results_Pr}.
    }
    \label{tab:results_Nd}
\end{table}

\begin{table}[tb]
    \centering
    \begin{tabular}{llrrccccccc}
\multicolumn{2}{c}{Tb emission line} & Energy & Photons & & & Stat. &  \multicolumn{3}{c}{\dotfill Systematic \dotfill} & Comb. \\
IUPAC & Sieg. & (eV) & counted & DE & SA & uncert. & $\delta$DE & $\delta$SA & $\delta$LS & uncert.  \\ \hline
    L$_3$M$_1$ & L$\ell$ & 5551.4 & 92900 & 0.183 & 0.523 & 0.004 & 0.017 & 0.057 & 0.01 & 0.061 \\
    L$_3$M$_4$ & L$\alpha_2$ & 6239.5 & 391100 & 0.202 & 0.595 & 0.002 & 0.001 & 0.003 &  & 0.004 \\
    L$_3$M$_5$ & L$\alpha_1$ & 6274.4 & 2916000 & 0.203 & 0.599 & 0.001 & --- & --- &  & 0.001 \\
    L$_3$N$_1$ & L$\beta_6$ & 7117.1 & 40600 & 0.200 & 0.672 & 0.014 & 0.012 & 0.059 & 0.25 & 0.258 \\
    n/d  & L$\beta_{14}$ & 7339.7 & 123600 & 0.197 & 0.689 & 0.009 & 0.015 & 0.073 & 0.10 & 0.125 \\
    L$_3$N$_\mathrm{4,5}$ & L$\beta_{2,15}$ & 7366.2 & 614000 & 0.196 & 0.691 & 0.003 & 0.015 & 0.075 & 0.02 & 0.079 \\
    L$_3$O$_1$ & L$\beta_7$ & 7467.6 & 8500 & 0.194 & 0.698 & 0.024 & 0.016 & 0.081 & 0.50 & 0.507 \\
    L$_3$N$_\mathrm{6,7}$ & Lu & 7512.4 & 7400 & 0.193 & 0.702 & 0.054 & 0.017 & 0.083 &  & 0.101 \\
\hline
    L$_2$M$_1$ & L$\eta$ & 6290.5 & 30500 & 0.203 & 0.598 & 0.027 & 0.010 & 0.048 &  & 0.057 \\
    L$_2$M$_4$ & L$\beta_1$ & 6977.2 & 1655000 & 0.202 & 0.659 & 0.001 & --- & --- & 0.01 & 0.010 \\
    n/d  & L$\beta'$ & 7003.4 & 35200 & 0.201 & 0.661 & 0.015 & 0.000 & 0.002 & 0.10 & 0.101 \\
    L$_2$N$_1$ & L$\gamma_5$ & 7856.8 & 8000 & 0.186 & 0.495 & 0.017 & 0.009 & 0.126 &  & 0.127 \\
    n/d  & L$\gamma_9$ & 8076.4 & 85400 & 0.181 & 0.512 & 0.005 & 0.010 & 0.113 &  & 0.114 \\
    L$_2$N$_4$ & L$\gamma_1$ & 8098.6 & 156100 & 0.180 & 0.514 & 0.003 & 0.011 & 0.112 &  & 0.112 \\
    L$_2$O$_1$ & L$\gamma_8$ & 8204.3 & 2320 & 0.178 & 0.523 & 0.051 & 0.011 & 0.106 &  & 0.118 \\
    L$_2$N$_\mathrm{6,7}$ & Lv & 8251.1 & 1310 & 0.177 & 0.526 & 0.052 & 0.012 & 0.103 &  & 0.116 \\
\hline
    L$_1$M$_2$ & L$\beta_4$ & 6942.1 & 133800 & 0.202 & 0.630 & 0.006 & 0.002 & 0.010 & 0.10 & 0.101 \\
    L$_1$M$_3$ & L$\beta_3$ & 7097.1 & 173300 & 0.200 & 0.643 & 0.005 & --- & --- & 0.06 & 0.060 \\
    L$_1$M$_4$ & L$\beta_{10}$ & 7431.0 & 3000 & 0.195 & 0.669 & 0.098 & 0.004 & 0.021 &  & 0.101 \\
    L$_1$M$_5$ & L$\beta_{9}$ & 7467.6 & 8500 & 0.194 & 0.671 & 0.024 & 0.004 & 0.023 & 0.50 & 0.501 \\
    L$_1$N$_2$ & L$\gamma_2$ & 8398.5 & 32900 & 0.173 & 0.433 & 0.017 & 0.011 & 0.157 & 0.80 & 0.815 \\
    L$_1$N$_3$ & L$\gamma_3$ & 8425.2 & 19600 & 0.172 & 0.435 & 0.009 & 0.011 & 0.155 &  & 0.156 \\
    L$_1$N$_\mathrm{4,5}$ & L$\gamma_{11}$ & 8557.3 & 450 & 0.169 & 0.444 & 0.118 & 0.012 & 0.149 &  & 0.190 \\
    L$_1$O$_\mathrm{2,3}$ & L$\gamma_{4,4'}$ & 8684.2 & 7170 & 0.166 & 0.453 & 0.017 & 0.013 & 0.143 &  & 0.144 \\
\end{tabular}

    \caption{Fluorescent-intensity measurements for the Tb foil sample, 250\,$\mu$m thick, 99.9\,\% pure. Columns are as in Table~\ref{tab:results_Pr}.
    }
    \label{tab:results_Tb}
\end{table}

\begin{table}[tb]
    \centering
    \begin{tabular}{llrrccccccc}
\multicolumn{2}{c}{Ho emission line} & Energy & Photons & & & Stat. &  \multicolumn{3}{c}{\dotfill Systematic \dotfill} & Comb. \\
IUPAC & Sieg. & (eV) & counted & DE & SA & uncert. & $\delta$DE & $\delta$SA & $\delta$LS & uncert.  \\ \hline
    L$_3$M$_1$ & L$\ell$ & 5940.2 & 113300 & 0.197 & 0.520 & 0.004 & 0.014 & 0.058 & 0.00 & 0.060 \\
    L$_3$M$_4$ & L$\alpha_2$ & 6678.5 & 392700 & 0.204 & 0.593 & 0.003 & 0.001 & 0.003 &  & 0.004 \\
    L$_3$M$_5$ & L$\alpha_1$ & 6719.7 & 3121000 & 0.204 & 0.596 & 0.001 & --- & --- &  & 0.001 \\
    L$_3$N$_1$ & L$\beta_6$ & 7639.5 & 46500 & 0.191 & 0.671 & 0.018 & 0.010 & 0.060 & 0.40 & 0.405 \\
    n/d  & L$\beta_{14}$ & 7884.3 & 87600 & 0.186 & 0.689 & 0.008 & 0.012 & 0.075 & 0.45 & 0.456 \\
    L$_3$N$_\mathrm{4,5}$ & L$\beta_{2,15}$ & 7910.3 & 645800 & 0.185 & 0.690 & 0.002 & 0.013 & 0.076 & 0.06 & 0.098 \\
    L$_3$O$_1$ & L$\beta_7$ & 8023.0 & 12700 & 0.182 & 0.698 & 0.018 & 0.013 & 0.083 &  & 0.086 \\
    L$_3$N$_\mathrm{6,7}$ & Lu & 8068.2 & 1890 & 0.181 & 0.701 & 0.044 & 0.014 & 0.085 &  & 0.097 \\
\hline
    L$_2$M$_1$ & L$\eta$ & 6786.5 & 26800 & 0.203 & 0.601 & 0.018 & 0.008 & 0.049 &  & 0.052 \\
    L$_2$M$_4$ & L$\beta_1$ & 7525.6 & 1120600 & 0.193 & 0.662 & 0.001 & --- & --- &  & 0.001 \\
    n/d  & L$\beta'$ & 7558.9 & 19000 & 0.192 & 0.664 & 0.020 & 0.000 & 0.002 &  & 0.020 \\
    L$_2$N$_1$ & L$\gamma_5$ & 8487.0 & 5380 & 0.171 & 0.502 & 0.024 & 0.007 & 0.123 &  & 0.126 \\
    n/d  & L$\gamma_9$ & 8731.4 & 66200 & 0.165 & 0.520 & 0.007 & 0.008 & 0.110 & 0.25 & 0.273 \\
    L$_2$N$_4$ & L$\gamma_1$ & 8750.8 & 90600 & 0.164 & 0.522 & 0.005 & 0.009 & 0.109 & 0.18 & 0.211 \\
    L$_2$O$_1$ & L$\gamma_8$ & 8866.1 & 1890 & 0.161 & 0.530 & 0.066 & 0.009 & 0.103 &  & 0.122 \\
    L$_2$N$_\mathrm{6,7}$ & Lv & 8915.5 & 770 & 0.160 & 0.534 & 0.074 & 0.009 & 0.100 &  & 0.125 \\
\hline
    L$_1$M$_2$ & L$\beta_4$ & 7471.2 & 156900 & 0.194 & 0.647 & 0.005 & 0.002 & 0.011 &  & 0.012 \\
    L$_1$M$_3$ & L$\beta_3$ & 7651.7 & 135900 & 0.191 & 0.661 & 0.007 & --- & --- & 0.14 & 0.140 \\
    L$_1$M$_4$ & L$\beta_{10}$ & 8002.7 & 3070 & 0.183 & 0.686 & 0.047 & 0.003 & 0.021 &  & 0.051 \\
    L$_1$M$_5$ & L$\beta_{9}$ & 8044.7 & 5520 & 0.182 & 0.688 & 0.026 & 0.003 & 0.023 &  & 0.035 \\
    L$_1$N$_2$ & L$\gamma_2$ & 9044.3 & 17800 & 0.157 & 0.454 & 0.013 & 0.009 & 0.157 & 0.20 & 0.255 \\
    L$_1$N$_3$ & L$\gamma_3$ & 9086.6 & 17100 & 0.155 & 0.457 & 0.010 & 0.009 & 0.155 & 0.20 & 0.253 \\
    L$_1$N$_\mathrm{4,5}$ & L$\gamma_{11}$ & 9229.6 & 250 & 0.152 & 0.467 & 0.148 & 0.010 & 0.148 &  & 0.210 \\
    L$_1$O$_\mathrm{2,3}$ & L$\gamma_{4,4'}$ & 9368.6 & 5260 & 0.148 & 0.476 & 0.017 & 0.010 & 0.142 &  & 0.143 \\
\end{tabular}

    \caption{Fluorescent-intensity measurements for the Ho foil sample, 300\,$\mu$m thick, 99.9\,\% pure. Columns are as in Table~\ref{tab:results_Pr}.
    }
    \label{tab:results_Ho}
\end{table}

This online supplement contains the data set from which the relative fluorescence intensities were derived: the raw spectra, the fitted intensities, the correction factors, and the uncertainties.

The file {\tt spectrum\_supplement.csv} contains the complete measured x-ray spectra recorded for each rare-earth sample from 1.5\,keV to 11\,keV. There is one bin per eV\@. The five columns are described in row one. They include the minimum energy of each bin, and the number of photons detected in each bin for each of the four samples. The spectral data set has been published previously as a supplement to \href{https://doi.org/10.1088/1681-7575/abd28a}{doi:10.1088/1681-7575/abd28a}.

Tables~\ref{tab:results_Pr}, \ref{tab:results_Nd}, \ref{tab:results_Tb}, and \ref{tab:results_Ho} contain the complete results for each element studied. The measured intensity of a line is the estimate based on a fit between the observed emission spectrum and a model consisting of one or more Voigt functions plus a locally linear background. The uncertainty is that reported by the fit and can be larger than a simple Poisson uncertainty, particularly for the lines of lower intensity, because of the non-zero background level. The detection efficiency (Section~\ref{sec:efficiency}) gives the values shown in Figure~\ref{fig:efficiency} and uncertainties as shown in Figure~\ref{fig:efficiency_uncert}.  The self-absorption factor is the fraction of emitted photons that also escape the emitting foil, as estimated by integrating Equation~\ref{eq:sa_weights} over the excitation spectrum (described in Section~\ref{sec:self-absorb}), along with uncertainties.
Both efficiency uncertainties are fractional uncertainties in the \emph{ratio} of the given quantity at the line in question to its value at the appropriate reference line. In the case of certain lines, an additional uncertainty $\delta$LS (``line separation'') arises because photons in two or more unresolved lines cannot be unambiguously allocated to the underlying transitions. The tables give as the total uncertainty the quadrature combination of the two efficiency uncertainties, the line-identification uncertainty, and the statistical uncertainty in the measured number of photons. The RFI is derived from these tables as the measured value divided by the product of the DE and SA efficiencies. All results for one subshell are rescaled by the value that fixes RFI=1 for the appropriate reference line (L$\alpha_1$, L$\beta_1$, or L$\beta_3$).

\begin{table}[tbh]
    \centering
    \begin{tabular}{llllllll}
     & Line & Element & Salem & McCrary & Elam & {\tt xraylib}  & This work \\ \hline
    L1 & L$\gamma_{2,3}$/L$\beta_3$ & Ho & &  .44(6) & .480 & .422 & .45(9) \\
    L1 & L$\gamma_{4,4'}$/L$\beta_3$ & Ho & &  .071(8) & & .055 & .069(14) \\ \hline
    L2 & L$\gamma_1$/L$\beta_1$ & Pr & &  & .156 & .168 & .174(28) \\
    & & Nd & .173(22) & & .159 & .169 & .158(28) \\
    & & Tb & .163(20) & & .171 & .175 & .135(15) \\
    & & Ho & .171(22) & .174(13)& .175 & .176 & .121(25) \\ \hline
    L3 & L$\alpha_2$/L$\alpha_1$ & Pr & .110(6) & & .111 & .113 & .119(1) \\
    & & Nd & .104(5) & & .111 & .113 & .114(1) \\
    & & Tb & .115(6) & & .111 & .113 & .135(1) \\
    & & Ho & .116(6) & & .111 & .113 & .127(1) \\ \hline
    L3 & L$\beta_{2,15}$/L$\alpha_1$ & Pr & .204(25) & & .211 & .182 & .253(19) \\
    & & Nd & .225(28) & & .212 & .183 & .250(18) \\
    & & Tb & .198(25) & & .208 & .188 & .189(15) \\
    & & Ho & .174(21) & .177(15) & .203 & .189 & .197(19) \\
    \end{tabular}
    \caption{Intensity ratios for three pairs of same-subshell emission lines, including diffraction-based measurements of Salem~\cite{Salem1971} and McCrary~\cite{McCrary1972}, theory values of Elam~\cite{Elam2002} and {\tt xraylib}/Scofield~\cite{Scofield1974,Schoonjans2011}, and this work.}
    \label{tab:salem_compare}
\end{table}

Table~\ref{tab:salem_compare} contains intensity ratios found in the literature for specific lines, as discussed in Section~\ref{sec:comparison}.

\end{document}